\documentclass[12pt]{article}
\usepackage{amsmath}
\usepackage{epsfig}
\usepackage{amssymb}
\usepackage{hyperref}
\usepackage{bbold}
\usepackage{braket}
\usepackage{dsfont}

\usepackage{mathtools}
\usepackage{amsfonts}
\usepackage[numbers,sort&compress]{natbib}
\allowdisplaybreaks

\def\be{\begin{equation}}
\def\ee{\end{equation}}
\def\beq{\begin{equation}}
\def\eeq{\end{equation}}
\def\beqa{\begin{eqnarray}}
\def\eeqa{\end{eqnarray}}
\def\ba{\begin{eqnarray}}
\def\ea{\end{eqnarray}}
\def\bea{\begin{eqnarray}}
\def\eea{\end{eqnarray}}

\newcommand\f[2]{\frac{#1}{#2}}

\def\beq{\begin{equation}}
\def\eeq{\end{equation}}
\def\beeq{\begin{eqnarray}}
\def\eeeq{\end{eqnarray}}
\def\to{\rightarrow}
\def\nn{\nonumber}

\def\b0{b_0}

\def\b0{b_0}

\begin{document}

\begin{titlepage}
\renewcommand{\thefootnote}{\fnsymbol{footnote}}
\begin{flushright}
%YITP-SB-18-35 \\
     \end{flushright}
\par \vspace{10mm}
\begin{center}
{\large \bf
Threshold resummation at N$^{3}$LL accuracy and \\[2mm] approximate N$^{3}$LO corrections to semi-inclusive DIS}\\

\vspace{8mm}

\today
\end{center}

\par \vspace{2mm}
\begin{center}
{\bf Maurizio Abele${}^{\,a}$,}
\hskip .2cm
{\bf Daniel de Florian${}^{\,b}$,}
\hskip .2cm
{\bf Werner Vogelsang${}^{\,a}$  }\\[5mm]
\vspace{5mm}
${}^{a}\,$ Institute for Theoretical Physics, T\"ubingen University, 
Auf der Morgenstelle 14, \\ 72076 T\"ubingen, Germany\\[2mm]
${}^b$ International Center for Advanced Studies (ICAS) and ICIFI, ECyT-UNSAM,\\
Campus Miguelete, 25 de Mayo y Francia, (1650) Buenos Aires, Argentina \\
\end{center}

%%%%%%%%%%%%%%%%%%%%%%%%%%%%%%%%%%%%%%%%%%%%%%%%%%%%%%%%
%%%%%%%%%%%%%%%%%%%%%%%%%%%       ABSTRACT      %%%%%%%%%%%%%%%%%%%
%%%%%%%%%%%%%%%%%%%%%%%%%%%%%%%%%%%%%%%%%%%%%%%%%%%%%%%%

\vspace{9mm}
\begin{center} {\large \bf Abstract} \end{center}
We advance the threshold resummation formalism for semi-inclusive deep-inelastic scattering (SIDIS)
to next-to-next-to-next-to-leading logarithmic (N$^{3}$LL) order, including the three-loop hard factor.
We expand the results in the strong coupling to obtain approximate next-to-next-to-next-to-leading order (N$^{3}$LO) 
corrections for the SIDIS cross section. In Mellin moment space, these corrections include 
all terms that are logarithmically enhanced at threshold, or that are constant.
%These corrections include all double distributions (that is, ``plus'' distributions and $\delta$-functions) in the partonic counterparts of 
%the customary SIDIS variables $x$ and $z$.  
We also consider a set of corrections that are suppressed near threshold.
Our numerical estimates show modest changes of the cross section by the approximate N$^{3}$LO terms, suggesting a very good 
perturbative stability of the SIDIS process.

\end{titlepage}  

\setcounter{footnote}{2}
\renewcommand{\thefootnote}{\fnsymbol{footnote}}

%%%%%%%%%%%%%%%%%%%%%%
\section{Introduction \label{intro}}
%%%%%%%%%%%%%%%%%%%%%%

The semi-inclusive deep-inelastic scattering (SIDIS) process $\ell p\to \ell hX$ has become a widely used 
probe of hadronic structure and hadronization phenomena. Its main uses are extractions of (polarized) parton 
distribution and fragmentation functions or combinations 
thereof~\cite{deFlorian:2009vb,Leader:2010rb,deFlorian:2014xna,Ethier:2017zbq,Bertone:2017tyb,Moffat:2021dji,Khalek:2021gxf,Abdolmaleki:2021yjf,Borsa:2021ran,Borsa:2022vvp}. 
In global analyses of these quantities SIDIS data can add useful information on, for example, the flavor structure of the sea quarks.
The future Electron Ion Collider (EIC) will allow precise measurements of SIDIS observables over wide kinematic regimes~\cite{Aschenauer:2019kzf}.

In a recent paper~\cite{Abele:2021nyo}, we have studied higher-order QCD corrections to the SIDIS cross section. Our approach 
was to use the threshold resummation formalism for SIDIS and carry out fixed-order expansions of the resummed cross section.
Threshold resummation for SIDIS was originally discussed in Ref.~\cite{Cacciari:2001cw} and then further developed in more
general terms in~\cite{Anderle:2012rq} and~\cite{Anderle:2013lka}. These papers formulated the resummation at next-to-leading 
logarithmic (NLL) accuracy. In~\cite{Abele:2021nyo} we extended the resummation to next-to-next-to-leading 
logarithm (NNLL), which also allowed us to obtain approximate fixed-order corrections to the hard scattering cross section for SIDIS
at next-to-next-to-leading order (NNLO) level. These results were used recently to obtain the first NNLO set of fragmentation
functions fit ``globally'' to SIDIS and electron-positron annihilation data~\cite{Borsa:2022vvp}. 

The purpose of the present paper is to advance our previous study to N$^{3}$LL and to again use the resummed cross section to
derive approximate fixed-order corrections to the SIDIS cross section, in this case at N$^{3}$LO. 
Our main motivation for this analysis is to further improve the perturbative
framework for SIDIS and to set the stage for precision analyses of SIDIS data from the future EIC in terms of 
parton distributions or fragmentation functions at high perturbative order.
While such analyses at N$^{3}$LO may presently still seem far off, the study of the perturbative stability of the SIDIS cross
section and its associated threshold 
resummation is in any case valuable. This becomes indeed possible by going to N$^{3}$LL and N$^{3}$LO and carrying out 
comparisons with lower orders. We also note that in our previous paper~\cite{Abele:2021nyo} we presented phenomenological
results only for the fixed-order (NNLO) corrections. Here we wish to carry out numerical studies also for the resummed
case, which provides another motivation for this study. 

In Sec.~\ref{pSIDIS} we give an overview of the kinematics of the process, introducing Mellin moments. 
Section~\ref{thresh} describes the threshold resummation framework. Section~\ref{hardfactor} is 
dedicated to the derivation of the three-loop hard factor to be used for obtaining N$^{3}$LL or N$^{3}$LO
results. In Sec.~\ref{N3LOexp} we carry out the expansion of the resummed results to N$^{3}$LO. 
Finally, Section~\ref{Pheno} presents some numerical studies in the EIC kinematical regime.

\section{Perturbative SIDIS cross section \label{pSIDIS}}

We consider the semi-inclusive deep-inelastic scattering (SIDIS) process
$\ell(k)\, p(P)\rightarrow \ell'(k')\, h(P_{h}) \, X$ with the momentum transfer $q = k-k'$.
It is described by the variables
\beeq\label{kindef}
Q^{2} 
	&= &-q^{2} = - (k-k')^{2}\; , \nn\\[2mm]
x 
	&= &\frac{Q^{2}}{2 P \cdot q} \; ,\nn\\[2mm]
y
	&= &\frac{P\cdot q}{P \cdot k}\; , \nn\\[2mm]
z
	&= &\frac{P \cdot P_{h}}{P \cdot q}\; .
\eeeq
We have $Q^{2} = xys$, with $\sqrt{s}$ the center-of-mass energy of the incoming electron and proton.
We follow Ref.~\cite{Anderle:2012rq} to write the spin-averaged SIDIS cross section as
\beq\label{sidiscrsec}
\frac{d^{3} \sigma^{h}}{dxdydz}\,=\,\frac{4 \pi \alpha^{2}}{Q^2} \left[ 
		\frac{1+(1-y)^{2}}{2y} \mathcal{F}^{h}_{T}(x,z,Q^2)
		+ \frac{1-y}{y} \mathcal{F}^{h}_{L}(x,z,Q^2)\right]\,,
\eeq
where $\alpha$ is the fine structure constant and ${\cal F}_T^h\equiv 2F_1^h$ 
and ${\cal F}_L^h\equiv F_L^h/x$ are the transverse and longitudinal structure 
functions. In what follows we will only treat the transverse structure function in the $q\rightarrow q$ or $\bar{q}\rightarrow \bar{q}$ channels,
which is the only channel that appears already at the lowest order (LO) of perturbation theory. We write
all equations for the spin-averaged case, although they will equally apply to the helicity-dependent one~\cite{Anderle:2013lka,Abele:2021nyo}.

Using factorization, the unpolarized structure functions may be written as double convolutions. For example, for
the transverse one we have
\begin{equation}
\label{F1hallorders}
\mathcal{F}^{h}_{T}(x,z,Q^2)
	\,=\, \sum_{f,f'} \int_{x}^{1} \frac{d\hat{x}}{\hat{x}} \int_{z}^{1} \frac{d \hat{z}}{\hat{z}}\,
	D^{h}_{f'}\left( \frac{z}{\hat{z}},\mu_{F}\right)
	\omega^{T}_{f'f} \left(\hat{x},\hat{z},\alpha_{s}(\mu_{R}), \frac{\mu_{R}}{Q}, \frac{\mu_{F}}{Q} \right)
	f\left( \frac{x}{\hat{x}},\mu_{F}\right)\,.
\end{equation}
Here $f(\xi,\mu_F)$ is the distribution of parton $f=q,\bar{q},g$ in the nucleon
at momentum fraction $\xi$ and factorization scale $\mu_F$, while $D^h_{f'} \left(\zeta,\mu_F\right)$
is the corresponding fragmentation function for parton $f'$ going to the
observed hadron $h$. For simplicity, the factorization scales are chosen to be equal in the initial and final state.
$\mu_R$ is the renormalization scale entering also the strong coupling $\alpha_s$. 
The functions $\omega^T_{f'f}$ are the transverse spin-averaged hard-scattering
coefficient functions which can be computed in QCD perturbation theory. Their expansions read
\beq\label{Cpert}
\omega^T_{f'f}\,=\,\omega^{T,(0)}_{f'f}+\frac{\alpha_s(\mu_R)}{\pi}\,\omega^{T,(1)}_{f'f}
+\left(\frac{\alpha_s(\mu_R)}{\pi}\right)^2 \omega^{T,(2)}_{f'f}+\left(\frac{\alpha_s(\mu_R)}{\pi}\right)^3 \omega^{T,(3)}_{f'f}+
{\cal O}(\alpha_s^4)\,.
\eeq
At LO we have for the $q\to q$ and $\bar{q}\to \bar{q}$ channels
\ba\label{LO}
\omega^{T,(0)}_{qq}(\hat{x},\hat{z})\,&=& e_q^2\,
\delta(1-\hat{x})\delta(1-\hat{z})\,,
\ea
with the quark's fractional charge $e_q$. The well known first-order coefficient function $\omega^{T,(1)}_{f'f}$ 
is for example available in~\cite{deFlorian:1997zj,Anderle:2012rq}.

In the following, it is convenient to take double Mellin moments of the SIDIS cross section, 
for which the convolutions in Eq.~(\ref{F1hallorders}) turn into ordinary products. We define
\ba\label{moms}
\tilde{{\cal F}}^h_T(N,M,Q^2)&\equiv&\int_0^1 dx\, x^{N-1}\int_0^1 dz\, z^{M-1}\,
{\cal F}^h_T(x,z,Q^2)\nn\\[2mm]
&=&\sum_{f,f'}\tilde{D}_{f'}^{h}(M,\mu_F)\,
\tilde{\omega}^T_{f'f}\left(N,M,\alpha_{s}(\mu_{R}), \frac{\mu_{R}}{Q}, \frac{\mu_{F}}{Q}\right)\tilde{f}(N,\mu_F)\,,
\ea
where
\ba
&&\tilde{f}(N,\mu_F)\equiv\int_0^1 dx \,x^{N-1}f(x,\mu_F),\nn\\[2mm]
&&\tilde{D}_{f'}^{h}(M,\mu_F)\equiv\int_0^1 dz \,z^{M-1}D^h_{f'}(z,\mu_F),\nn\\[2mm]
&&\tilde{\omega}^T_{f'f}\left(N,M,\alpha_{s}(\mu_{R}), \frac{\mu_{R}}{Q}, \frac{\mu_{F}}{Q}\right)\,\equiv\,
\int_0^1 d\hat{x}\,\hat{x}^{N-1}\int_0^1 d\hat{z}\,\hat{z}^{M-1}\,
\omega^T_{f'f}\left(\hat{x},\hat{z},\alpha_{s}(\mu_{R}), \frac{\mu_{R}}{Q}, \frac{\mu_{F}}{Q}\right) \,.\hspace*{6mm}
\ea
As a result the structure functions can be obtained from 
the moments of the parton distribution functions and fragmentation
functions, and the double-Mellin moments of  the partonic hard-scattering functions.

For the perturbative expansion given in Eq.~(\ref{Cpert}) we have in moment space 
at lowest order according to Eq.~(\ref{LO})
\ba
\tilde{\omega}^{T,(0)}_{qq}(N,M)\,&=&e_q^2 .
\ea
The corresponding moments of the next-to-leading order (NLO) terms $\omega^{T,(1)}_{f'f}$
may be found in Refs.~\cite{deFlorian:1997zj,Anderle:2012rq}. In the following, we consider logarithmic 
higher-order corrections to the hard-scattering functions that arise at large values of $\hat{x}$ and $\hat{z}$
or, equivalently, at large $N$ and $M$.

\section{Threshold resummation at N$^3$LL accuracy \label{thresh}}

The resummation of threshold logarithms for SIDIS 
was extensively studied in Refs.~\cite{Cacciari:2001cw,Sterman:2006hu,Anderle:2012rq,Abele:2021nyo}. The NNLL 
resummation formula for the unpolarized SIDIS transverse structure function was 
discussed in Ref.~\cite{Abele:2021nyo}. The resummed partonic transverse structure function
takes the form
\beeq\label{SIDISres}
\tilde{\omega}^{T,{\mathrm{res}}}_{qq}\left( N,M,\alpha_{s}(\mu_{R}), \frac{\mu_{R}}{Q}, \frac{\mu_{F}}{Q}\right)&=&
e_q^2  \,H^{\mathrm{SIDIS}}_{qq}  \left(\alpha_{s}\big(\mu_{R}\big),\frac{\mu_{R}}{Q}, \frac{\mu_{F}}{Q}
\right) \,	\widehat{C}_{qq} \left(\alpha_{s}\big(\mu_{R}\big),\frac{\mu_{R}}{Q}\right) \nn\\[2mm]
&\times&\exp\left\{\int_{Q^{2}/(\bar{N}\bar{M})}^{Q^2} 
	\frac{d \mu^{2}}{\mu^{2}} \left[  A_{q}\big(\alpha_{s}(\mu)\big) \ln\left( \frac{\mu^{2} \bar{N}\bar{M}}{Q^{2}}\right)
		- \frac{1}{2} \widehat{D}_{q} \big(\alpha_{s} (\mu)\big)\right]\right.\nn\\[2mm]
	&+&\left. \ln\bar{N} \int_{Q^2}^{ \mu^{2}_{F}} \frac{d \mu^{2}}{\mu^{2}} A_{q}\big(\alpha_{s}(\mu)\big)
	\,+\,\ln\bar{M} \int_{Q^2}^{ \mu^{2}_{F}} \frac{d \mu^{2}}{\mu^{2}} A_{q}\big(\alpha_{s}(\mu)\big)\right\}\,,
\eeeq
which actually holds to any logarithmic order. As stated earlier, our goal is to set up the formalism for
resummation to N$^3$LL. In Eq.~(\ref{SIDISres}) we have
\beq\label{Nbar}  
\bar{N}\,=\,N\,{\mathrm{e}}^{\gamma_E}\quad\mathrm{and}\quad\bar{M}\,=\,M\,{\mathrm{e}}^{\gamma_E}\,,
\eeq
with the Euler constant $\gamma_E$. Each of the functions $A_q,\widehat{D}_q,H_{qq}^{\mathrm{SIDIS}},\widehat{C}_{qq}$
is a perturbative series in the strong coupling. We write the corresponding expansions generically as
\beq\label{Qexp}
{\cal Q}\,=\,\sum_{k=0}^\infty  \left(\frac{\alpha_s(\mu_R)}{\pi}\right)^k\,{\cal Q}^{(k)}\,,
\eeq
where ${\cal Q}=A_q,\widehat{D}_q,H_{qq}^{\mathrm{SIDIS}},\widehat{C}_{qq}$. 
We note that $A_q^{(0)}=\widehat{D}_q^{(0)}=\widehat{D}_q^{(1)}=0$. 
To achieve N$^3$LL accuracy, we need $A_q$ to order $\alpha_s^4$ and all other functions to order $\alpha_s^3$.
The corresponding coefficients are collected in Appendix~\ref{sec:appendixAndim}. The main new 
ingredient not directly known from the literature is the $\bar{N}, \bar{M}$-independent coefficient
$H_{qq}^{\mathrm{SIDIS},(3)}$ whose derivation will be presented below in Sec.~\ref{hardfactor}.
The other prefactor $\widehat{C}_{qq}$ in Eq.~(\ref{SIDISres}) collects all moment-independent terms of
the resummed exponent; see~\cite{Catani:2003zt,Hinderer:2018nkb,Abele:2021nyo}. The formulas needed for
its derivation to order $\alpha_s^3$ may be found in Ref.~\cite{Catani:2003zt}.

In order to explicitly obtain the structure function resummed to N$^3$LL we now expand the exponents in 
Eq.~(\ref{SIDISres}) appropriately. The operations are quite standard. We obtain 
\beeq\label{SIDISres2}
&&\hspace*{-1.2cm}
\tilde{\omega}^{T,\mathrm{res}}_{qq}\left( N,M,\alpha_{s}(\mu_{R}), \frac{\mu_{R}}{Q}, \frac{\mu_{F}}{Q}\right)\,=\,
e_q^2  \,H^{\mathrm{SIDIS}}_{qq} \left(\alpha_{s}(\mu_R), \frac{\mu_{R}}{Q},\frac{\mu_{F}}{Q} \right)\,\widehat{C}_{qq}
\left(\alpha_{s}(\mu_R), \frac{\mu_{R}}{Q} \right)\nn\\[2mm]
&\times& \exp\left\{\frac{\lambda_{NM}}{2 b_{0} \alpha_{s}(\mu_{R})}\,h^{(1)}_{q}\left(\frac{\lambda_{NM}}{2}\right)+
h^{(2)}_{q}\left(\frac{\lambda_{NM}}{2}, \frac{\mu_{R}}{Q}, \frac{\mu_{F}}{Q}\right) \right. \nn \\[2mm]
&&\qquad\;\;+ \left. \alpha_{s}(\mu_{R})\; h^{(3)}_{q}\left(\frac{\lambda_{NM}}{2}, \frac{\mu_{R}}{Q}, \frac{\mu_{F}}{Q}\right)+\alpha^2_{s}(\mu_{R})\; h^{(4)}_{q}\left(\frac{\lambda_{NM}}{2}, \frac{\mu_{R}}{Q}, \frac{\mu_{F}}{Q}\right) \right\}\,,
\eeeq
where
\beeq\label{lambda}
\lambda_{NM} \,\equiv\, b_{0} \,\alpha_{s}(\mu_{R}) \big(\ln\bar{N} + \ln \bar{M} \big)\,.
\eeeq
The functions $h^{(k)}_{q}$ impart resummation to N$^{k-1}$LL accuracy. The first three are well known in
the literature:
\beeq\label{n3llexp}
	 h^{(1)}_{q}\left(\lambda\right)
	 	&=&  \frac{A^{(1)}_{q}}{\pi b_{0} \lambda} \left[ 2\lambda + (1 - 2\lambda) \ln(1-2\lambda) \right]\,,\nn\\[2mm]
	 h^{(2)}_{q}\left(\lambda\right)
	 	&= & -\frac{A^{(2)}_{q}}{\pi^2 b^{2}_{0}} \left[ 2\lambda +\ln(1-2\lambda) \right]\nonumber\\[2mm]
	 	&+&\frac{A^{(1)}_{q} b_{1}}{\pi b^{3}_{0} }\left[ 2\lambda + \ln(1-2\lambda) + \frac{1}{2}\ln^{2}(1-2\lambda) \right]\nn\\[2mm]
	 	&+& \frac{A^{(1)}_{q}}{\pi b_{0}} 2\lambda \ln\frac{\mu^{2}_{F}}{Q^{2}}
	 	- \frac{A^{(1)}_{q}}{\pi b_{0}}\left[ 2\lambda + \ln(1-2\lambda)  \right] \ln\frac{\mu^{2}_{R}}{Q^{2}}\,,\nonumber\\[2mm]
	h^{(3)}_{q}\left(\lambda\right)
	&=&-\frac{A^{(2)}_{q} b_{1}}{\pi^{2} b^{3}_{0}}\frac{1}{1-2\lambda}[2\lambda + \ln(1-2\lambda) + 2\lambda^{2}]\nonumber\\[2mm]
	&+&\frac{A^{(1)}_{q}b^{2}_{1}}{\pi b^{4}_{0} (1-2\lambda)}\left[ 2\lambda^{2} +2\lambda \ln(1-2\lambda) +\frac{1}{2} \ln^{2}(1-2\lambda) \right] \nonumber\\[2mm]
	&+&\frac{A^{(1)}_{q} b_{2}}{\pi b^{3}_{0}}\left[ 2\lambda + \ln(1-2\lambda) + \frac{2\lambda^{2}}{1-2\lambda} \right]
	+\frac{2A^{(3)}_{q}}{\pi^{3}b^{2}_{0}}\frac{\lambda^{2}}{1-2\lambda}\nonumber\\[2mm]
	&+& \frac{2 A^{(2)}_{q}}{\pi^{2}b_{0}}\lambda \ln\frac{\mu^{2}_{F}}{Q^{2}}
	-\frac{A^{(1)}_{q}}{\pi} \lambda \ln^{2}\frac{\mu^{2}_{F}}{Q^{2}}
	+\frac{2 A^{(1)}_{q}}{\pi}  \lambda \ln\frac{\mu^{2}_{F}}{Q^{2}}\ln\frac{\mu^{2}_{R}}{Q^{2}}\nn\\[2mm]
	&-&\frac{1}{1-2\lambda}\left( \frac{A^{(1)}_{q} b_{1}}{\pi b^{2}_{0}}[2\lambda + \ln(1-2\lambda)] - 
	\frac{4 A^{(2)}_{q} }{\pi^{2} b_{0}} \lambda^{2}\right)\ln\frac{\mu^{2}_{R}}{Q^{2}}\nonumber\\[2mm]
	&+&\frac{2 A^{(1)}_{q}}{\pi}\frac{\lambda^{2}}{1-2\lambda}\ln^{2}\frac{\mu^{2}_{R}}{Q^{2}}
	-\frac{\widehat{D}^{(2)}_{q}}{\pi^{2}b_{0}}\frac{\lambda}{1-2\lambda}\,.
\eeeq
The function $h^{(4)}_{q}$, needed for N$^3$LL resummation, is found to be
\beeq
h^{(4)}_{q}(\lambda)
	&=& \frac{1}{(1 -2\lambda)^2}\left(  \frac{A^{(2)}_{q} b^{2}_1}{ \pi^2 b_0^4 }\left[-\frac{8 }{3 }\lambda^3-\lambda^2+\lambda+\frac{1}{2 }\ln ^2(1-2 \lambda)+\frac{1}{2 }\ln (1-2 \lambda)\right] \right.\nn\\[2mm]
	&+& \frac{A^{(2)}_{q} b_2}{\pi^2 b_0^3 } \frac{8 }{3}\lambda^3 + \frac{A^{(1)}_{q} b^3_1}{ \pi  b_0^5}\left[\frac{8 }{3}\lambda^3+ 2 \lambda^2 \ln (1-2 \lambda)-\frac{1}{6}\ln^3(1-2 \lambda)\right] \nn\\[2mm]
	&+& \frac{A^{(1)}_{q} b_1 b_2}{\pi  b_0^4 }\left[-\frac{16 }{3 }\lambda^3+3 \lambda^2-\lambda-4 \lambda^2 \ln (1-2 \lambda)+2 \lambda \ln (1-2 \lambda)-\frac{1}{2 }\ln (1-2 \lambda)\right] \nn\\[2mm]
	&+& \frac{A^{(1)}_{q} b_3}{ \pi  b_0^3}\left[\frac{8 }{3 }\lambda^3-3 \lambda^2+\lambda+2 \lambda^2 \ln (1-2 \lambda)-2 \lambda \ln (1-2 \lambda)+\frac{1}{2 }\ln (1-2 \lambda)\right] \nn\\[2mm]
	&+& \frac{A^{(3)}_{q}b_1}{ \pi^3 b_0^3}\left[ \frac{8 }{3 }\lambda^3-\lambda^2-\lambda-\frac{1}{2}\ln (1-2 \lambda) \right] + \frac{A^{(4)}_{q}}{\pi^4 b_0^2}\left[2 \lambda^2 - \frac{8}{3} \lambda^3\right] \nn\\[2mm]
	&+&\left. \frac{\widehat{D}^{(2)}_{q}b_1}{ \pi^2 b_0^2}\left[\lambda- \lambda^2 +\frac{1}{2}\ln (1-2 \lambda)\right] + \frac{\widehat{D}^{(3)}_{q}}{\pi^3 b_0}\left[\lambda^2-\lambda\right] \right)\,.
\eeeq
This result is in agreement with that given in Ref.~\cite{Moch:2005ba} for the Drell-Yan process.
For simplicity, we have set the renormalization and factorization scales to $Q$. 
The results presented in Eq.~(\ref{SIDISres2}) may be used to obtain N$^{3}$LO (that is,
${\cal O}(\alpha_s^3)$) expansions of the hard-scattering function $\tilde{\omega}^T_{qq}$. 
This expansion will be carried out in Sec.~\ref{N3LOexp}. 

We stress that all terms generated by Eq.~(\ref{SIDISres2}) are either logarithmic or constant near threshold.
The full hard-scattering function in Mellin space will, at any order in perturbation theory, also contain terms that are suppressed
by powers of $1/N$ and/or $1/M$. Such terms are often referred to as ``next-to-leading power (NLP)'' corrections. 
As discussed in our previous paper~\cite{Abele:2021nyo} (see also references therein), one can straightforwardly account
for the dominant NLP terms by multiplying the resummed cross section in Eq.~(\ref{SIDISres}) by the two factors
\beq\label{nlpfactors}
\exp\left\{-\int^{Q^2/(\bar{N}\bar{M})}_{ \mu^{2}_{F}} \frac{d \mu^{2}}{\mu^{2}}
\frac{\alpha_{s}(\mu)}{\pi}\, \frac{C_F}{2N}\right\}\,
\exp\left\{-\int^{Q^2/(\bar{N}\bar{M})}_{ \mu^{2}_{F}} \frac{d \mu^{2}}{\mu^{2}}
\frac{\alpha_{s}(\mu)}{\pi}\, \frac{C_F}{2M}\right\}\,.
\eeq
where the coefficients $-C_F/(2N)$ and $-C_F/(2M)$ in the exponents correspond to the NLP terms in the 
LO diagonal evolution kernels for the quark parton distributions and the quark fragmentation functions, respectively. 
At N$^3$LO the two exponential factors, when combined with the resummed exponents
in Eq.~(\ref{SIDISres}), will generate all terms of the form $\alpha_s^3 \ln^n(N)\ln^m(M)(1/N+1/M)$, with
$n+m=5$. 

\section{The hard factor at three loops \label{hardfactor}}

The factor $H^{\mathrm{SIDIS}}_{qq}$ is derived from the finite part of the virtual corrections to the process 
$\gamma^*q\to q$. The basic ingredient is the renormalized spacelike form quark factor, from which one needs to 
subtract the infrared divergencies via a suitable method developed in Refs.~\cite{Catani:2013tia,Catani:2014uta}. 
For our present purposes, we will need the renormalized three-loop form factor, which was derived 
in~\cite{Baikov:2009bg,Gehrmann:2010ue,Lee:2010cga}\footnote{
We note that recently even the four-loop results were published~\cite{Lee:2022nhh}.} 
and reads in dimensional regularization with $d=4-2\epsilon$ space-time dimensions:
\begin{equation}
	F_{q}(q^{2})
		\,=\, F^{(0)}_{q} +  \frac{\alpha_{s}}{\pi} \, F^{(1)}_{q}
			+ \left( \frac{\alpha_{s}}{\pi}  \right)^{2} F^{(2)}_{q}
			+ \left( \frac{\alpha_{s}}{\pi}  \right)^{3} F^{(3)}_{q}
			+ \mathcal{O}(\alpha_{s}^{4})\,,
\end{equation}
where 
\beeq
F^{(0)}_{q} &= &1 \,,\nn \\[2mm]
F^{(1)}_{q}
	&=&C_{F}\left[-\frac{1}{2 \epsilon^2} 
		-\frac{3}{4 \epsilon }
		+\frac{\pi ^2}{24}-2
		+\left(\frac{7 \zeta (3)}{6}+\frac{\pi ^2}{16}-4\right) \epsilon  +\left(\frac{7 \zeta (3)}{4}+\frac{47 \pi ^4}{2880}+\frac{\pi ^2}{6}-8\right) \epsilon^2
		\right.\nn\\[2mm]
	&+&\left.
	\left(
	\frac{31 \zeta (5)}{10} + \frac{14 \zeta (3)}{3} - \frac{7 \pi^2 \zeta (3)}{72}
	+\frac{47 \pi ^4}{1920}+\frac{\pi ^2}{3}-16
	\right) \epsilon^3
			\right.\nn\\[2mm]
	&+&\left.
	\left(
	   \frac{28 \zeta (3)}{3}-\frac{7 \pi ^2 \zeta (3)}{48}-\frac{49 \zeta
    (3)^2}{36}+\frac{93 \zeta (5)}{20}+\frac{2 \pi ^2}{3}+\frac{47 \pi
    ^4}{720}+\frac{949 \pi ^6}{241920}-32
	\right) \epsilon^4
	+\mathcal{O}\left(\epsilon ^5\right)
	\right]\,,\nn\\[2mm]
F^{(2)}_{q}&=&C^{2}_{F}
\left[\frac{1}{8 \epsilon ^4}+\frac{3}{8 \epsilon ^3}+\left(\frac{41}{32}-\frac{\pi ^2}{48}\right)\frac{1}{\epsilon^2}+\left(\frac{221}{64}-\frac{4 \zeta (3)}{3}\right)\frac{1}{\epsilon} 
\right.\nn\\[2mm]
&-&\left. \frac{29 \zeta (3)}{8}-\frac{13 \pi ^4}{576}+\frac{17 \pi ^2}{192}+\frac{1151}{128}\right]\nn \\[2mm]
&+& C_{F} C_{A} \left[ \frac{11}{32 \epsilon ^3}+\left( \frac{1}{9}+\frac{\pi ^2}{96} \right)\frac{1}{\epsilon^2}+\left(\frac{13 \zeta (3)}{16}-\frac{11 \pi ^2}{192}-\frac{961}{1728}\right)\frac{1}{\epsilon }  \right.\nn \\[2mm]
&+&\left.\frac{313 \zeta (3)}{144}+\frac{11 \pi ^4}{720}-\frac{337 \pi ^2}{1728}-\frac{51157}{10368}\right] \nn\\[2mm]
&+& C_{F} N_{f}  \left[-\frac{1}{16 \epsilon ^3}-\frac{1}{36 \epsilon^2}+\left( \frac{65}{864}+\frac{\pi ^2}{96} \right)\frac{1}{\epsilon }+
\frac{\zeta (3)}{72}+\frac{23 \pi ^2}{864}+\frac{4085}{5184} \right]
\nn\\[2mm]
&+& \left(  C_F^2 \left[  -\frac{839 \zeta (3)}{48}+\frac{7 \pi ^2 \zeta (3)}{18}-\frac{23 \zeta
    (5)}{10}+\frac{5741}{256}+\frac{71 \pi ^2}{128}-\frac{19 \pi ^4}{320} \right] \right.\nn \\[2mm]
&+&\left. C_F C_A \left[ 
 \frac{5893 \zeta (3)}{432}-\frac{89 \pi ^2 \zeta (3)}{288}+\frac{51 \zeta
    (5)}{16}-\frac{1319701}{62208}-\frac{8089 \pi ^2}{10368}+\frac{229 \pi
    ^4}{5760}
     \right] \right.\nn \\[2mm]
&+&\left. C_F N_f \left[  -\frac{119 \zeta (3)}{216}+\frac{108653}{31104}+\frac{497 \pi
    ^2}{5184}+\frac{\pi ^4}{2880}  \right]
     \right) \epsilon \nn\\[2mm]
&+& \left( 
C_F^2 \left[ -\frac{6989 \zeta (3)}{96}+\frac{9 \pi ^2 \zeta (3)}{16}+\frac{163 \zeta
    (3)^2}{9}-\frac{231 \zeta (5)}{40}+\frac{27911}{512}+\frac{613 \pi
    ^2}{256}-\frac{3401 \pi ^4}{11520}+\frac{223 \pi ^6}{17280}
 \right] \right.\nn \\[2mm]
&+&\left. C_F C_A \left[   
\frac{148861 \zeta (3)}{2592}-\frac{10 \pi ^2 \zeta (3)}{27}-\frac{569 \zeta
    (3)^2}{48}+\frac{2809 \zeta (5)}{240}-\frac{28437757}{373248}-\frac{165205
    \pi ^2}{62208}
\right. \right.\nn \\[2mm]
&+&\left. \left.
\frac{48127 \pi ^4}{207360}-\frac{809 \pi ^6}{241920}
 \right] 
    + C_F N_f \left[ 
  -\frac{3581 \zeta (3)}{1296}-\frac{5 \pi ^2 \zeta (3)}{108}-\frac{59 \zeta
    (5)}{120}+\frac{2379989}{186624}  
    \right. \right.\nn \\[2mm]
&+&\left. \left.
 \frac{9269 \pi ^2}{31104}-\frac{145 \pi
    ^4}{20736} 
  \right]
  \right) \epsilon^2
+\mathcal{O}\left(\epsilon ^3\right)\,,\nn \\[2mm]
% F^{(3)}_{q} &= &1 \,,\nn \\[2mm]
F^{(3)}_{q}
	&=&C_{F}^3 \Bigg[ -\frac{1}{48\epsilon^6} -\frac{3}{32\epsilon^5} + \frac{1}{\epsilon^4} \left(\frac{\pi ^2}{192}-\frac{25}{64} \right)
	+ \frac{1}{\epsilon^3} \left( \frac{25 \zeta (3)}{48}-\frac{83}{64}-\frac{\pi ^2}{128}\right)\nn \\[2mm] &&
	+ \frac{1}{\epsilon^2} \left(\frac{69 \zeta (3)}{32}-\frac{515}{128}-\frac{77 \pi ^2}{768}+\frac{71 \pi
    ^4}{7680} \right)
	+ \frac{1}{\epsilon} \left(\frac{2119 \zeta (3)}{192}-\frac{107 \pi ^2 \zeta (3)}{576}+\frac{161 \zeta
    (5)}{80}-\frac{9073}{768} \right. \nn \\[2mm] &&
    \left.
    -\frac{467 \pi ^2}{768}+\frac{487 \pi ^4}{15360} \right)
    + \frac{2669 \zeta (3)}{64}+\frac{61 \pi ^2 \zeta (3)}{384}-\frac{913 \zeta
    (3)^2}{96}+\frac{2119 \zeta (5)}{160}-\frac{53675}{1536}\nn \\[2mm]
     && -\frac{13001 \pi
    ^2}{4608}+\frac{12743 \pi ^4}{92160}-\frac{9095 \pi ^6}{3483648}
	\Bigg] \nn \\[2mm]
	&&+C_{F}^2 C_A \Bigg[  -\frac{11}{64\epsilon^5} 
	+ \frac{1}{\epsilon^4} \left( -\frac{361}{1152}-\frac{\pi ^2}{192} \right)
	+ \frac{1}{\epsilon^3} \left( -\frac{13 \zeta (3)}{32}-\frac{1703}{3456}+\frac{9 \pi ^2}{256} \right)\nn \\[2mm]
	&&+ \frac{1}{\epsilon^2} \left( -\frac{241 \zeta (3)}{288}+\frac{1705}{1296}+\frac{1487 \pi
    ^2}{13824}-\frac{83 \pi ^4}{11520} \right)
       + \frac{1}{\epsilon} \left( -\frac{4151 \zeta (3)}{384}+\frac{215 \pi ^2 \zeta (3)}{1152}-\frac{71 \zeta
    (5)}{32} \right. \nn \\[2mm] &&
    \left.
+\frac{374149}{31104}+\frac{31891 \pi ^2}{41472}-\frac{2975 \pi
    ^4}{165888} \right)
     -\frac{19933 \zeta (3)}{384}-\frac{403 \pi ^2 \zeta (3)}{576}+\frac{101 \zeta
    (3)^2}{12}-\frac{3445 \zeta (5)}{288} \nn \\[2mm] &&
    +\frac{11169211}{186624}+\frac{537803
    \pi ^2}{124416}-\frac{723739 \pi ^4}{4976640}-\frac{18619 \pi ^6}{17418240}
	\Bigg] \nn \\[2mm]
	&&+C_{F}^2 N_f \Bigg[  \frac{1}{32\epsilon^5} +\frac{35}{576\epsilon^4} 
	+ \frac{1}{\epsilon^3} \left( \frac{139}{1728}-\frac{\pi ^2}{128} \right)	
	+ \frac{1}{\epsilon^2} \left(  -\frac{55 \zeta (3)}{288}-\frac{775}{5184}-\frac{133 \pi ^2}{6912} \right)	
	\nn \\[2mm]
     &&+ \frac{1}{\epsilon} \left(  \frac{469 \zeta (3)}{1728}-\frac{24761}{15552}-\frac{2183 \pi
    ^2}{20736}-\frac{287 \pi ^4}{82944} \right)
    + \frac{21179 \zeta (3)}{5184}+\frac{35 \pi ^2 \zeta (3)}{1152}-\frac{193 \zeta
    (5)}{288}\nn \\[2mm]
     && -\frac{691883}{93312}-\frac{16745 \pi ^2}{31104}-\frac{8503 \pi
    ^4}{2488320}	
		\Bigg] \nn \\[2mm]
		&&+C_{F} C_A^2  \Bigg[  -\frac{1331}{5184 \epsilon^4} 
	+ \frac{1}{\epsilon^3} \left( \frac{1433}{7776}-\frac{55 \pi ^2}{5184}\right)
	+ \frac{1}{\epsilon^2} \left( -\frac{451 \zeta (3)}{864}+\frac{11669}{31104}+\frac{1625 \pi
    ^2}{31104}-\frac{11 \pi ^4}{12960}\right)\nn \\[2mm]
	&&
	+ \frac{1}{\epsilon} \left( \frac{1763 \zeta (3)}{864}-\frac{11 \pi ^2 \zeta (3)}{432}-\frac{17 \zeta
    (5)}{24}-\frac{139345}{559872}-\frac{7163 \pi ^2}{93312}-\frac{83 \pi
    ^4}{17280}\right)
	+\frac{505087 \zeta (3)}{31104}\nn \\[2mm]
	&&
	+\frac{13 \pi ^2 \zeta (3)}{36}-\frac{71 \zeta
    (3)^2}{36}-\frac{217 \zeta (5)}{288}-\frac{51082685}{3359232}-\frac{412315
    \pi ^2}{279936}+\frac{22157 \pi ^4}{622080}-\frac{769 \pi ^6}{326592}
	\Bigg] \nn \\[2mm]
		&&+C_{F} N_f^2  \Bigg[  -\frac{11 }{ 1296\epsilon^4} 
	- \frac{1}{1944 \epsilon^3} 
	+ \frac{1}{\epsilon^2} \left(\frac{23}{2592}+\frac{\pi ^2}{864} \right)
	+ \frac{1}{\epsilon} \left(-\frac{\zeta (3)}{648}+\frac{2417}{139968}-\frac{5 \pi ^2}{2592} \right)
	\nn \\[2mm]
	&&-\frac{13 \zeta (3)}{486}-\frac{190931}{839808}-\frac{103 \pi
    ^2}{3888}-\frac{47 \pi ^4}{77760}
	\Bigg] \nn \\[2mm]
		&&+C_{F} C_A N_f  \Bigg[  \frac{ 121}{1296 \epsilon^4} 
	+ \frac{1}{\epsilon^3} \left( \frac{5 \pi ^2}{2592}-\frac{47}{972}\right)
	+ \frac{1}{\epsilon^2} \left( \frac{53 \zeta (3)}{432}-\frac{517}{3888}-\frac{119 \pi ^2}{7776}\right)
	\nn \\[2mm]
	&&
	+ \frac{1}{\epsilon} \left( -\frac{241 \zeta (3)}{1296}-\frac{8659}{139968}+\frac{1297 \pi
    ^2}{46656}+\frac{11 \pi ^4}{8640}\right)
	-\frac{67 \zeta (3)}{27}+\frac{\pi ^2 \zeta (3)}{288}-\frac{\zeta
    (5)}{48}+\frac{1700171}{419904}\nn \\[2mm]
	&&+\frac{115555 \pi ^2}{279936}+\frac{\pi
    ^4}{31104}
	\Bigg] \nn \\[2mm]	
		&&+ C_{F} N_{f,V} \left(\frac{C_{A}^{2}-4}{ 2 C_{A}}\right)\left( \frac{7 \zeta (3)}{48}-\frac{5 \zeta (5)}{6}+\frac{5 \pi ^2}{96}+\frac{1}{8}-\frac{\pi ^4}{2880} \right)+\mathcal{O}\left(\epsilon\right)\,.
\eeeq
Here we have kept terms of higher order  in $\epsilon$ in the one-loop and two-loop results since these turn
out to make finite contributions in the end. In the above expressions, $\zeta(j)$ is the Riemann zeta function, 
$N_f$ is the number of flavors, and $C_F=4/3,C_A=3$. 
For purely electromagnetic interactions the factor $N_{f,V=\gamma}$ becomes \cite{Gehrmann:2010ue}
\begin{equation}
N_{f,\gamma} = \frac{\sum_{q}e_q}{e_q}\; .
\end{equation}
As shown in Refs.~\cite{Catani:2013tia,Catani:2014uta}, the hard coefficient may be extracted from the form factor in the following way. 
Adapted to the case of SIDIS we have from~\cite{Catani:2014uta}
\begin{equation}
H_{qq}^{\mathrm{SIDIS}} \big(\alpha_{s}(Q)\big)\,=\, 
\left|\, 	\big[ 1 - \tilde{I}_q\big(\epsilon, \alpha_s(Q)\big) \big] F_q\,\right|^2	\,, 
\label{eq:Cthdef}
\end{equation}
where $\tilde{I}_q$ is an operator that removes the poles of the form factor and makes the necessary
soft and collinear adjustments needed to extract the hard coefficient. It is given in~\cite{Catani:2014uta} in 
terms of a convenient all-order form:
\begin{equation}
1 - \tilde{I}_q (\epsilon, \alpha_s)\,=\, 
\exp\left\{ R_q\left(\epsilon, \alpha_{s}\right) - i \Phi_q\left(\epsilon,\alpha_{s}\right)\right\}\,,
\end{equation}
with functions $R_q$ and $\Phi_q$ that each are perturbative series. The phase $\Phi_q$ does not contribute
in our case since we take the absolute square in Eq.~(\ref{eq:Cthdef}). The function $R_q$ effects the
cancelation of infrared divergences from the quark form factor. It can be expressed in terms 
of a soft and a collinear part:
\begin{equation}
R_q(\epsilon, \alpha_{s})
	=  R^{\mathrm{\,soft}}_q(\epsilon, \alpha_{s}) + R^{\mathrm{\,coll}}_q(\epsilon, \alpha_{s})\,,
\end{equation}
where for N$^{3}$LL accuracy
\beeq
R_q^{\mathrm{\,soft}}(\epsilon, \alpha_{s}) 
	&=& C_F \left( \frac{\alpha_{s}}{\pi}R_q^{\mathrm{\,soft}\, (1)}(\epsilon) + \left(\frac{\alpha_{s}}{\pi}\right)^{2} 
	R_q^{\mathrm{\,soft}\, (2)}(\epsilon) +  \left(\frac{\alpha_{s}}{\pi}\right)^{3} 
	R_q^{\mathrm{\,soft}\, (3)}(\epsilon) +\mathcal{O}(\alpha^{4}_{s})  \right)\; ,\nn\\[2mm]
R^{\mathrm{\,coll}}_q(\epsilon, \alpha_{s})
	&=&\frac{\alpha_{s}}{\pi}R_q^{\mathrm{\,coll}\, (1)}(\epsilon) + \left(\frac{\alpha_{s}}{\pi}\right)^{2} 
	R_q^{\mathrm{\,coll}\, (2)}(\epsilon) + \left(\frac{\alpha_{s}}{\pi}\right)^{3} 
	R_q^{\mathrm{\,coll}\, (3)}(\epsilon) + \mathcal{O}(\alpha^{4}_{s})\,,
\eeeq
with
\beeq
R_q^{\mathrm{\,soft}\, (1)}(\epsilon)
	&=& \frac{1}{2 \epsilon^{2}} -\frac{\pi ^2}{8} \; ,\nn \\[2mm]
R_q^{\mathrm{\,soft}\, (2)}(\epsilon)
	&=& - \frac{3\pi b_0}{8\epsilon^{3}} + \frac{1}{8 \epsilon^{2}} \,\frac{A^{(2)}_{q}}{C_{F}} \nn\\[2mm]
	&-& \frac{1}{16 \epsilon} \left[ C_{A} \left(7 \zeta(3)+\frac{11 \pi ^2}{36}-\frac{202}{27}\right)+ N_{f}\left(\frac{28}{27}-\frac{\pi ^2}{18}\right)\right] \nn\\[2mm]
	&+&C_{A} \left(-\frac{187 \zeta(3)}{144}+\frac{\pi ^4}{288}-\frac{469 \pi ^2}{1728}+\frac{607}{648}\right)+ N_{f} \left(\frac{17 \zeta(3)}{72}+\frac{35 \pi ^2}{864}-\frac{41}{324}\right)	\; , \nn \\[2mm]	
R_q^{\mathrm{\,soft}\, (3)}(\epsilon)
	&=& \f{11 (b_0 \pi)^2}{36\epsilon^4} - 
	\f{2 b_1\,\pi^2}{9\epsilon^3}
	-\f{5}{36\epsilon^3}b_0\pi \,\frac{A^{(2)}_{q}}{C_{F}}
	+\f{1}{18\epsilon^2}\,\frac{A^{(3)}_{q}}{C_{F}}\; \nn \\[2mm]
	&& +\f{1}{24\epsilon^2}b_0\pi  \left[ C_{A} \left(7 \zeta(3)+\frac{11 \pi ^2}{36}-\frac{202}{27}\right)+ N_{f}\left(\frac{28}{27}-\frac{\pi ^2}{18}\right)\right] \; \nn \\[2mm]
	&&- \frac{1}{48\epsilon}\; \Bigg[  C_A^2\left(-\f{136781}{5832} + \f{6325}{1944}\pi^2 - \f{11}{45}\pi^4+
\f{329}{6}\zeta(3) - \f{11}{9}\pi^2\zeta(3) - 24\zeta(5)\right)\nn\\[2mm]
&&+C_A\,N_f\left(\f{5921}{2916} - \f{707}{972}\pi^2 + \f{\pi^4}{15} -
\f{91}{27}\zeta(3)\right)+C_F\,N_f\left(\f{1711}{216} - \f{\pi^2}{12} - \f{\pi^4}{45} - \f{38}{9}\zeta(3)\right)\nn\\
&&+N_f^2\left(\f{260}{729} + \f{5}{162}\pi^2 - \f{14}{27}\zeta(3)\right) \Bigg]
	\nn \\[2mm]
	&& + C_A^2\Bigg(\f{5211949}{1679616}-\f{578479}{559872}\pi^2+\f{9457}{311040}\pi^4+\f{19}{326592}\pi^6-\f{64483}{7776}\zeta(3)+\f{121}{192}\pi^2\zeta(3)\nn\\[2mm]
&&+\f{67}{72}\zeta(3)^2-\f{121}{144}\zeta(5)\Bigg)+C_A\, N_f \left(-\f{412765}{839808}+\f{75155}{279936}\pi^2-\f{79}{9720}\pi^4+\f{154}{81}\zeta(3)\right.\nn\\
&&\left. -\f{11}{288}\pi^2\zeta(3)-\f{1}{24}\zeta(5)\right)
+C_F\,N_f\left(-\f{42727}{62208}+\f{605}{6912}\pi^2+\f{19}{12960}\pi^4+\f{571}{1296}\zeta(3)\right.\nn\\[2mm]
&&\left.-\f{11}{144}\pi^2\zeta(3)+\f{7}{36}\zeta(5)\right)+N_f^2 \left(-\f{2}{6561} - \f{101}{7776}\pi^2 + \f{37}{77760}\pi^4 -
\f{185}{1944}\zeta(3)\right) ,
\eeeq
and
\beeq
R_q^{\mathrm{\,coll}\, (1)}(\epsilon)
	&= &\frac{3}{4 \epsilon} C_{F} \; ,\nn\\[2mm]
R_q^{\mathrm{\,coll}\, (2)}(\epsilon)
	&=& -\frac{3 \pi b_0}{8\epsilon^{2}}C_{F}+ \frac{1}{8\epsilon} \left[ C_{F}^2 \left(6 \zeta(3)-\frac{\pi ^2}{2}+\frac{3}{8}\right)
	+\, C_{A} C_{F} \left(-3 \zeta(3) +\frac{11 \pi ^2}{18}+\frac{17}{24}\right)  \right. \nonumber\\[2mm]
	&&\left. \hspace*{2.9cm}+\,C_{F} N_{f}\left(-\frac{1}{12}-\frac{\pi ^2}{9}\right)\right] . \nonumber\\[2mm]
R_q^{\mathrm{\,coll}\, (3)}(\epsilon)
		&=&  C_{F}  \frac{(b_0 \pi)^2}{4 \epsilon^3}  -C_{F}   \frac{b_1 \pi^2}{4 \epsilon^2}  
	- \frac{b_0 \pi}{12\epsilon^2}  \left[ C_{F}^2 \left(6 \zeta(3)-\frac{\pi ^2}{2}+\frac{3}{8}\right)
	 \right. \nonumber\\[2mm]
	&&\left. +\, C_{A} C_{F} \left(-3 \zeta(3) +\frac{11 \pi ^2}{18}+\frac{17}{24}\right) +\,C_{F} N_{f}\left(-\frac{1}{12}-\frac{\pi ^2}{9}\right)\right] \nonumber\\[2mm]
	&& +\frac{1}{24\epsilon} \Bigg[ 
	C_F^3\left(\f{29}{16}+\f{3}{8}\pi^2+\f{\pi^4}{5}+\f{17}{2}\zeta(3)-\f{2}{3}\pi^2\zeta(3)-30\zeta(5)\right)\nn\\[2mm]
&&+C_F^2C_A\left(\f{151}{32}-\f{205}{72}\pi^2-\f{247}{1080}\pi^4+\f{211}{6}\zeta(3)+\f{1}{3}\pi^2\zeta(3)+15\zeta(5)\right)\nn\\[2mm]
&&+C_A^2C_F\left(-\f{1657}{288}+\f{281}{81}\pi^2-\f{\pi^4}{144}-\f{194}{9}\zeta(3)+5\zeta(5)\right)\nn\\[2mm]
&&+C_F^2 N_f\left(-\f{23}{8}+\f{5}{36}\pi^2+\f{29}{540}\pi^4-\f{17}{3}\zeta(3)\right)+C_FN_f^2\left(-\f{17}{72}+\f{5}{81}\pi^2-\f{2}{9}\zeta(3)\right)\nn\\[2mm]
&&+C_FC_A N_f\left(\f{5}{2}-\f{167}{162}\pi^2+\f{\pi^4}{360}+\f{25}{9}\zeta(3)\right)  \Bigg] \, .
\eeeq
The coefficients  $b_{0}$ and  $b_{1}$ can be found in Appendix~\ref{sec:appendixAndim}. Inserting all terms into Eq.~(\ref{eq:Cthdef}) and expanding in $\alpha_s$, all poles in powers of $1/\epsilon$ cancel. The final expression for $H_{qq}^{\mathrm{SIDIS}}$ up to three loops can be found in Appendix~\ref{sec:appendixAndim}.

\section{Expansion to N$^3$LO \label{N3LOexp}}

We are now ready to present the N$^3$LO (${\cal O}(\alpha_s^3)$) expansion for the SIDIS $q\to q$ hard-scattering function near
threshold. To write our formulas compactly we introduce
\beq
{\cal L}\,\equiv\,\frac{1}{2}\left( \ln (\bar{N})  + \ln (\bar{M})\right) \,.
\eeq 
The coefficients $\tilde{\omega}^{T,(1)}_{qq}$ and $\tilde{\omega}^{T,(2)}_{qq}$ in Eq.~(\ref{Cpert}) were already given in our 
previous paper~\cite{Abele:2021nyo}; for completeness, we recall them in Appendix \ref{appendix:hardscat}. 
For the approximate N$^3$LO terms we find:
\beeq \label{eqnlo1b}
&&\hspace*{-1cm}\frac{1}{e_q^2}\,\tilde{\omega}^{T,(3)}_{qq}\left(N,M, 1, 1\right)\,=\,\frac{4 }{3} C_F^3\mathcal{L}^6 +  \frac{8}{3}C_{F}^2\pi b_{0}\mathcal{L}^5  + \mathcal{L}^4 \left[ C_{F}^3 \left(-8 + \frac{\pi ^2}{3}\right)  - \frac{11}{27} C_{F}C_{A}N_{f} \right.\nn\\[2mm]
&+&\,\left. C_{F}^2 C_{A}\left( \frac{67}{9}-\frac{\pi ^2}{3} \right)+ \frac{121}{108} C_{F}C^2_{A} - \frac{10}{9} C_{F}^2 N_{f} + \frac{1}{27} C_{F}N^{2}_{f}\right] \nn\\[2mm]
&+&\, \mathcal{L}^3 \left[ C_{F}C_{A}N_{f}\left(\frac{\pi ^2}{27}-\frac{289}{162}\right) + C_{F}^2 C_{A}\left(-7 \zeta (3)+\frac{11 \pi ^2}{54}+\frac{70}{27}  \right)\right. \nn\\[2mm]
&&\left. \,+\, C_{F}C^2_{A}\left( \frac{445}{81}-\frac{11 \pi ^2}{54} \right) + C_{F}^2 N_{f}\left( -\frac{\pi ^2}{27}-\frac{17}{54} \right) + \frac{10}{81} C_{F}N^{2}_{f}\right] \nn\\[2mm]
&+&\, \mathcal{L}^2 \left[ C_{F}^3 \left(\frac{511}{32} -\frac{15 }{2}\zeta (3)-\frac{\pi ^2}{8}-\frac{\pi ^4}{30}\right) + C_{F}C_{A}N_{f}\left(\frac{5 \pi ^2}{54}-\frac{2051}{648}\right)\right. \nn\\[2mm]
&&\,+\,  C_{F}^2 C_{A}\left( \frac{151 }{18}\zeta (3)+\frac{143 \pi ^2}{216}-\frac{\pi ^4}{120}-\frac{8893}{288}\right) + C_{F}^2 N_{f}\left( \frac{10 }{9}\zeta (3)+\frac{67}{18}-\frac{\pi ^2}{108} \right)  \nn\\[2mm]
&&\left.\, + \,   C_{F}C^2_{A}\left( \frac{11 \pi ^4}{360}-\frac{11 }{2}\zeta (3)-\frac{67 \pi ^2}{108}+\frac{15503}{1296} \right) + \frac{25}{162}C_{F}N^{2}_{f}\right] \nn\\[2mm]
&+& \mathcal{L} \left[ C_{F}^2 C_{A}\left(14 \zeta (3) -\frac{1}{12} 7 \pi ^2 \zeta (3)+\frac{101 \pi ^2}{162}-\frac{404}{27} \right) + C_{F} N^{2}_{f}\left( \frac{\zeta (3)}{9}+\frac{58}{729} \right)  \right.  \nn\\[2mm]
&&\, + \, C_{F} C^{2}_{A} \left( \frac{11 \pi ^2 }{36}\zeta (3)-\frac{1541 }{108}\zeta (3)+6 \zeta (5)-\frac{11 \pi ^4}{720}-\frac{799 \pi ^2}{1944}+\frac{297029}{23328} \right) \nn\\[2mm]
&&\,+\,\left.  C^{2}_{F} N_{f}\left( \frac{19 }{18}\zeta (3)+\frac{\pi ^4}{180}+\frac{3}{32}-\frac{7 \pi ^2}{81} \right) + C_{F} C_{A} N_{f} \left(\frac{113 }{108}\zeta (3)+\frac{103 \pi ^2}{1944}-\frac{\pi ^4}{360}-\frac{31313}{11664}\right) \right]\nn\\[2mm]
&+& C_{F}^3\left(\frac{\zeta (3)^2}{2}+\frac{5 \pi ^2 }{6}\zeta (3)-\frac{115 }{16}\zeta (3)+\frac{83 }{4}\zeta (5)+\frac{761 \pi ^6}{136080}+\frac{37 \pi ^4}{2880}-\frac{5599}{384}-\frac{1663 \pi ^2}{1152}\right) \nn\\[2mm]
&+& C_{F}^2 C_{A}\left(\frac{37 }{12}\zeta (3)^2-\frac{119 \pi ^2 }{72}\zeta (3)-\frac{12877 }{432}\zeta (3)-\frac{689 }{72}\zeta (5)+\frac{40223 \pi ^2}{10368}+\frac{74321}{2304}-\frac{149 \pi ^6}{27216}-\frac{1147 \pi ^4}{38880}\right) \nn\\[2mm]
&+& C^2_{F} N_{f}\left(\frac{1181 }{216}\zeta (3)-\frac{19 }{18}\zeta (5)-\frac{421}{192}-\frac{559 \pi ^2}{1296}-\frac{29 \pi ^4}{9720}\right) + C_{F} N^{2}_{f}\left(\frac{\zeta (3)}{324}-\frac{23 \pi ^2}{432}-\frac{7081}{15552}-\frac{17 \pi ^4}{19440}\right) \nn\\[2mm]
&+& C_{F} C_{A}^2\left(-\frac{25 }{12}\zeta (3)^2+\frac{569 \pi ^2 }{864}\zeta (3)+\frac{139345 }{5184}\zeta (3)-\frac{51 }{16}\zeta (5)+\frac{17 \pi ^6}{34020}+\frac{3103 \pi ^4}{311040}-\frac{93889 \pi ^2}{31104}\right. \nn\\[2mm]
&-&\left.\frac{1505881}{62208}\right) + C_{F} C_{A} N_{f} \left(\frac{\pi ^2 }{216}\zeta (3)-\frac{383 }{81}\zeta (3)-\frac{\zeta (5)}{8}+\frac{469 \pi ^4}{77760}+\frac{6493 \pi ^2}{7776}+\frac{110651}{15552}\right)\nn\\[2mm]
&+& C_F N_{f,V}\frac{\left(C^{2}_A-4\right) }{C_A}\left(\frac{7 \zeta (3)}{48}-\frac{5 \zeta (5)}{6}+\frac{5 \pi ^2}{96}+\frac{1}{8}-\frac{\pi ^4}{2880}\right)
\,+\,2\,C_F^3\, {\cal L}^5 \left(\frac{1}{N}+\frac{1}{M}\right) \,.
\eeeq
As before, we have set $\mu_R = \mu_F = Q$ for simplicity. We stress that the corrections given by this expression
include all terms that are logarithmically enhanced at threshold, or that are constant. In physical space these
are terms with double distributions (that is, ``plus'' distributions and $\delta$-functions) in $\hat{x}$ and $\hat{z}$.  

The last term in Eq.~(\ref{eqnlo1b}) represents the dominant NLP contributions. Note that upon expansion beyond NLO the exponential factors in~(\ref{nlpfactors}) 
will also generate terms with inverse powers $1/N^2, 1/M^2$ and higher, which we have discarded for consistency since they are far beyond
the approximations we make. We will see later that these terms are numerically very small.

\section{Phenomenological predictions \label{Pheno}}

We will now present some phenomenological predictions for the transverse SIDIS cross section at NNLL and N$^3$LL, as well as for the 
expansion to N$^{3}$LO. We will also compare to our previous NNLO results of~\cite{Abele:2021nyo}. 

In order to obtain results for the transverse structure function ${\cal F}^h_T(x,z,Q^2)$ in physical $x,z$ space 
we need to invert its Mellin moments $\tilde{{\cal F}}^h_T(N,M,Q^2)$ in Eq.~\eqref{moms}. This is achieved by the inverse double-Mellin transform
\begin{equation}\label{eq:inverse}
{\cal F}^h_T(x,z,Q^2)\,=\, \int_{{\cal C}_N}
\frac{d N}{2\pi i}\,x^{-N}  \int_{{\cal C}_M}
\frac{d M}{2\pi i}\,z^{-M}\,
\tilde{{\cal F}}^h_T(N,M,Q^2)\,,
\end{equation}
where ${\cal C}_N$ and ${\cal C}_M$ denote integration contours in the complex plane, one for each Mellin inverse. 
We adopt the minimal prescription of Ref. \cite{Catani:1996yz} to treat the Landau pole present in the resummed exponents in
Eqs.~(\ref{SIDISres2}),(\ref{n3llexp}) at $\lambda_{NM}=1$, or (see Eq.~(\ref{lambda})),
\beq
\bar{N} \bar{M}\,=\,{\mathrm{e}}^{1/(b_{0} \,\alpha_{s}(\mu_{R}))}\,.
\eeq
According to the minimal prescription, the two contours need to be chosen such that all singularities in the 
complex plane lie to their left, except for the Landau pole. We parameterize the two contours as
\begin{equation}\label{NMcon}
N = c_{N} \pm \zeta \mathrm{e}^{\pm i \phi_{N}}\; , \quad
M = c_{M} + \xi \mathrm{e}^{i \phi_{M}}\; ,
\end{equation} 
with $\zeta,\xi\in [0,\infty]$ as contour parameters, where $c_N=1.8$ and $c_M=3.3$. We furthermore choose $\phi_N=3\pi/4$;
the two signs for the $N$-contour in~(\ref{NMcon}) select the two branches in the complex plane. 
As $N$ moves along its contour, the position of the Landau pole relevant for the $M$-integral will move as well, mapping out
a trajectory in the plane. This implies that the angle $\phi_M$ needs to be chosen as a function of $N$, so that 
during the $M$ integration this trajectory is never crossed. A more detailed description of the inverse double-Mellin transform 
can be found in Ref.~\cite{Anderle:2012rq}.

We note that we only consider the transverse structure function in~(\ref{sidiscrsec}) here. The longitudinal one is
suppressed near threshold, even beyond the dominant NLP terms we have included for ${\cal F}^h_T$. While it
would be very interesting to also investigate higher-order corrections to ${\cal F}^h_L$, this is beyond the scope of
this work. In what follows, we also discard the contributions by the $q\to g$ and $g\to q$ channels to the structure
function. These are fully known only to NLO. We could include the contributions at NLO level in our approximate 
NNLO, N$^3$LO results to be presented below, but this would simply amount to a uniform shift of all results by a few
per cent, which is not really relevant for our main goal of analyzing the structure of higher-order contributions in the $q\to q$ channel.

For the parton distribution functions and fragmentation functions we choose the NNLO sets of Ref.~\cite{Hou:2019efy}
and Ref.~\cite{Anderle:2016czy}, respectively. Clearly, in order to present true N$^3$LL or N$^3$LO results, we 
would need PDFs and FFs evolved at those same orders, which are currently not yet available. We therefore
keep the renormalization and factorization scales at $\mu_R=\mu_F=Q$ and do not investigate the scale
dependence of our results. In this sense we use the NNLO parton distributions and fragmentation functions
as templates for the N$^3$LO ones, which should be adequate for a first analysis of the beyond-NNLO effects
we are interested in. We note that the scale dependence of the transverse SIDIS cross section 
was anyway found to be rather small already at NNLO in Ref.~\cite{Abele:2021nyo}. Note that
we ``match'' all results to NLO, so that the NLO corrections for the $q\to q$ channel are always 
included {\it exactly}. 
%We do not carry out any
%``matching'' of the resummed results to a full fixed-order result since the latter is only known at NLO, where
%it is known from~\cite{Abele:2021nyo} that the threshold terms reproduce the full NLO cross section extremely
%well. 

Our predictions will refer to the unpolarized $\ell p\to \ell\pi^+X$ process appropriate for the future EIC with $\sqrt{s} = 100$~GeV. 
We focus on the $z$-dependence of 
the cross section and integrate over $y \in[0.1,0.9]$ and $x \in[0.1,0.8]$. We choose $x$ and $z$ 
to be rather large so that we are safely in the threshold regime. Because of the relation $Q^2=xys$, 
our choice of kinematics implies $Q^2>100$~GeV$^2$ for the EIC. We furthermore require $W>7$~GeV, where $W^2=Q^2(1-x)/x+m_p^2$,
with $m_p$ the proton mass. 

\begin{figure}[t!]
\vspace*{-5mm}
\includegraphics[scale=0.9]{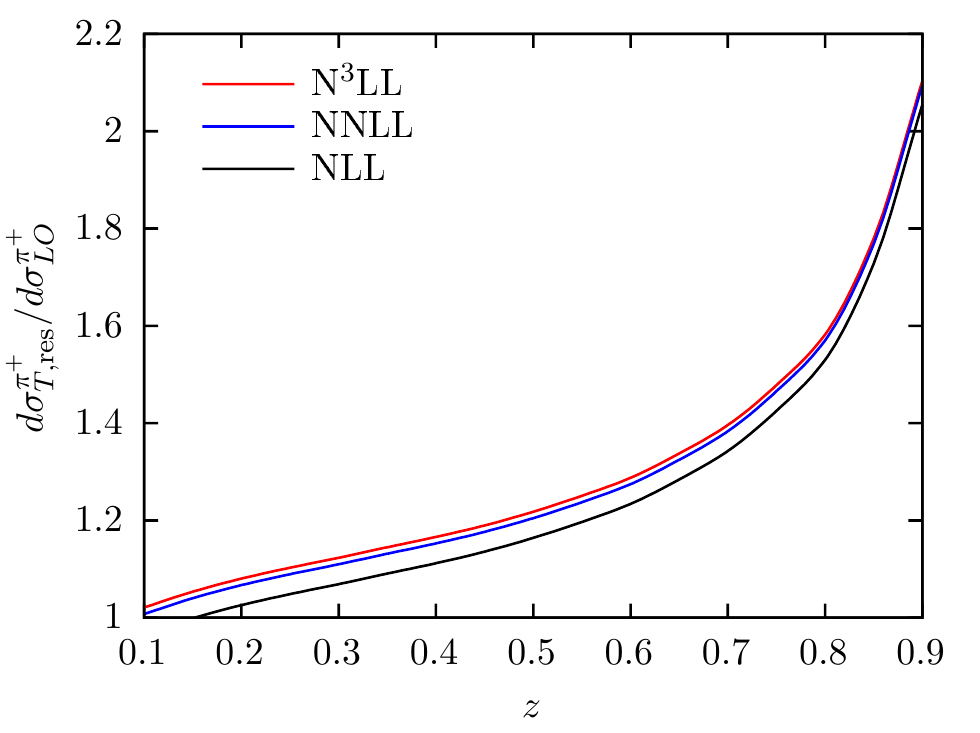}

\vspace*{-6.75cm}
\hspace*{8.6cm}
\includegraphics[scale=0.9]{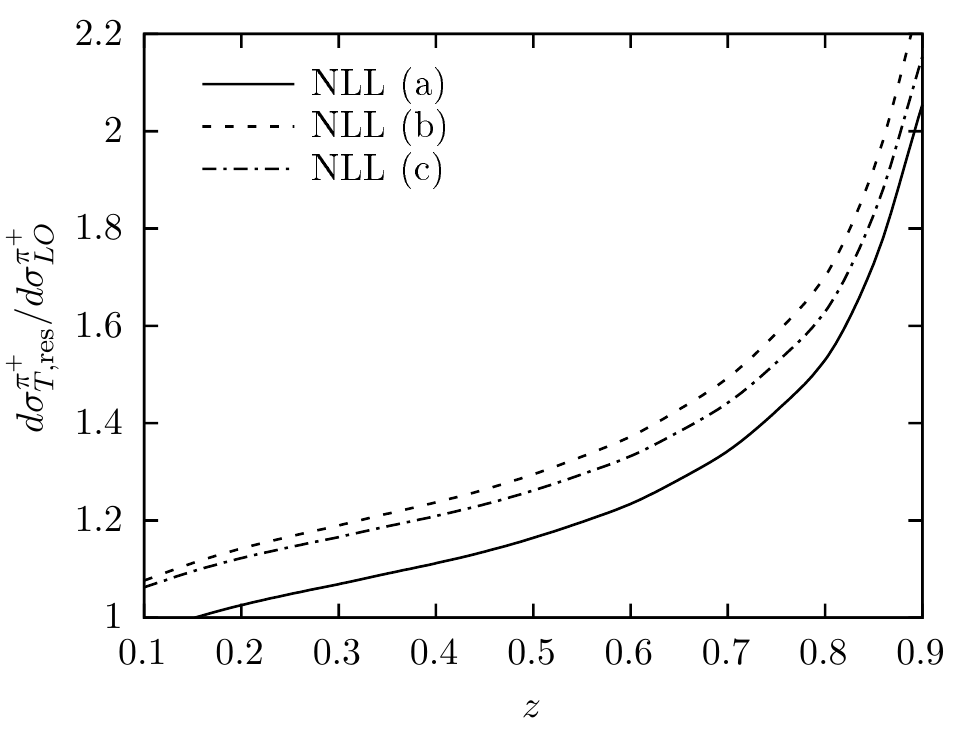}

\includegraphics[scale=0.9]{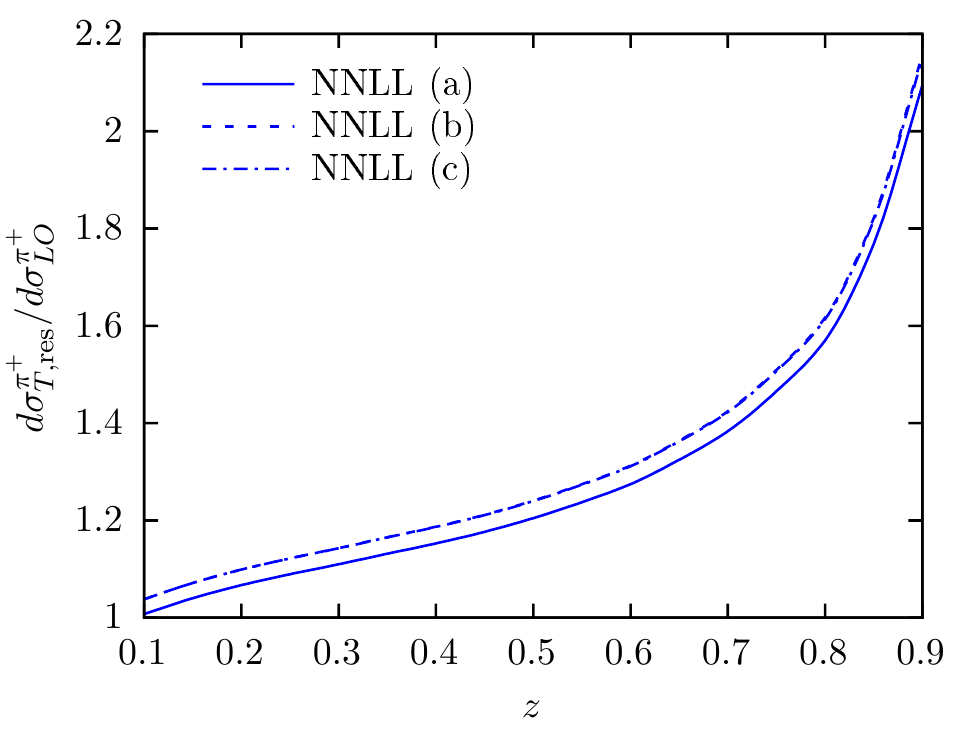}

\vspace*{-6.75cm}
\hspace*{8.6cm}
\includegraphics[scale=0.9]{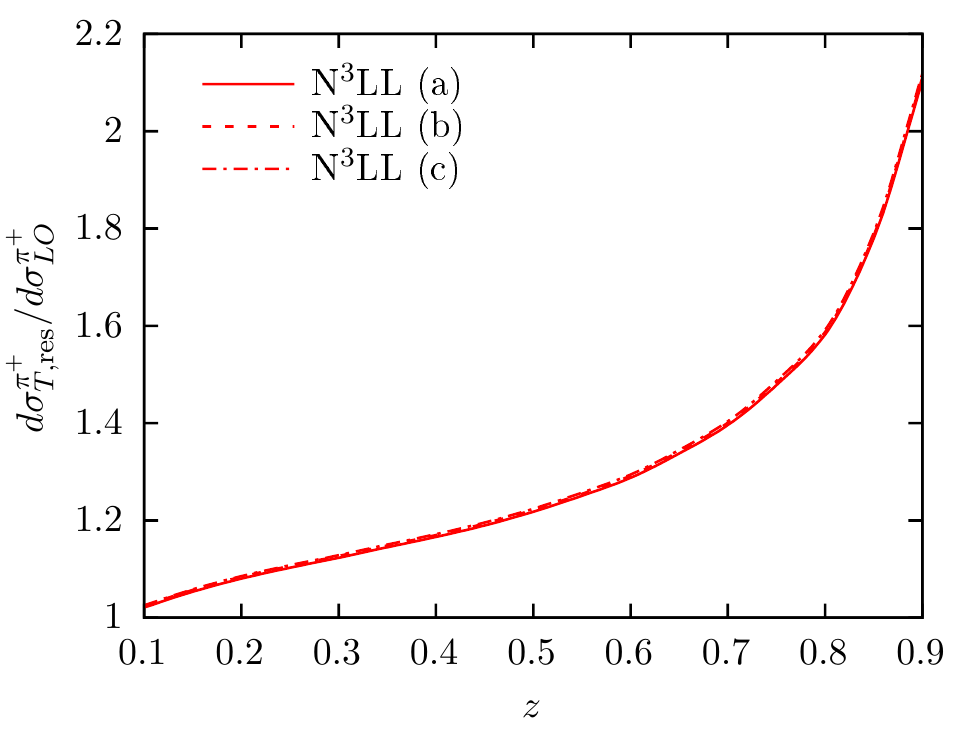}

\caption{{\it Ratios of resummed results for the unpolarized $\ell p\to \ell\pi^+X$ transverse cross section in the $q\to q$ channel to the LO one, 
for EIC kinematics with $x\in [0.1,0.8]$. Upper left: comparison of NLL (black), NNLL (blue), and $N^3$LL (red) resummation.
Upper right and lower panel: Results for resummation schemes (a),(b),(c) as described in the text at NLL, NNLL, and $N^3$LL.}
\label{fig1}}
\end{figure}

We begin by comparing fully resummed results obtained at various different levels of logarithmic accuracy. The upper left
part of Fig.~\ref{fig1} shows the NLL, NNLL, and N$^3$LL resummed cross sections as functions of $z$, 
normalized to the LO one. As one can see, the NNLL terms lead to a significant enhancement over NLL,
while the additional terms arising at N$^3$LL only lead to a modest further increase of the cross section. 
This result demonstrates that the resummed SIDIS cross section has excellent perturbative stability.

We can further investigate the improvements provided by going to NNLL and N$^3$LL. To this end, we note that even at a given logarithmic order 
the resummation formula in Eq.~(\ref{SIDISres}) may actually be used in various ways that are all equivalent in terms of their perturbative 
content, but differ numerically. Let us refer to the corresponding choices as {\it resummation schemes}.  We consider three such schemes:
\begin{description}
\item[Scheme (a)] Here we use Eq.~(\ref{SIDISres}) as written. That is, we keep the functions $H^{\mathrm{SIDIS}}_{qq}$ and $\widehat{C}_{qq}$ 
as separate factors, each its own perturbative series of the form~(\ref{Qexp}). Also, we use the Mellin moments $N$ and $M$ precisely in the
form $\bar{N}$ and $\bar{M}$ as defined in~(\ref{Nbar}). This scheme has been used for the first plot in Fig.~\ref{fig1}. 
\item[Scheme (b)] Here we expand the product $H^{\mathrm{SIDIS}}_{qq}\times\widehat{C}_{qq}$ in Eq.~(\ref{SIDISres}) strictly 
to the desired order. That is, suppose we are at NLL where $H^{\mathrm{SIDIS}}_{qq}=1+\frac{\alpha_s}{\pi}H^{\mathrm{SIDIS},(1)}_{qq}$
and $\widehat{C}_{qq}=1+\frac{\alpha_s}{\pi}\widehat{C}_{qq}^{(1)}$, then we use 
$H^{\mathrm{SIDIS}}_{qq}\times\widehat{C}_{qq}=1+\frac{\alpha_s}{\pi}(H^{\mathrm{SIDIS},(1)}_{qq}+\widehat{C}_{qq}^{(1)})$ and 
drop terms of ${\cal O}(\alpha_s^2)$. We continue to use the variables $\bar{N}$ and $\bar{M}$. 
\item[Scheme (c)] Here we first use the expansion of $H^{\mathrm{SIDIS}}_{qq}\times\widehat{C}_{qq}$ as for
scheme (b). In addition, we use~(\ref{Nbar}),(\ref{lambda}) to write
\beq\label{lambda2}
\lambda_{NM}\,=\,b_{0} \, \alpha_{s}(\mu_{R}) \big(\ln N + \ln  M \big) + 2 \gamma_E b_0 \alpha_s(\mu_R)\,.
\eeq
The terms with the Euler constant lead to modifications of the functions $h_q^{(k>1)}$ in Eq.~(\ref{n3llexp})~\cite{Catani:2003zt,Moch:2005ba}. They
evidently also generate non-logarithmic corrections in the resummed exponent. These may be expanded out perturbatively,
so that they migrate from the exponent to an $N,M$-independent prefactor. This prefactor is then expanded along with the 
factor $H^{\mathrm{SIDIS}}_{qq}\times\widehat{C}_{qq}$ into a {\it single} perturbative function that now multiplies the 
resummed exponent, the latter now being a function of $N$ and $M$ rather than of $\bar{N}$ and $\bar{M}$. 
\end{description}
It is immediately clear that the three resummation schemes are indeed equivalent for a given logarithmic accuracy. The remaining three
plots in Fig.~\ref{fig1} compare the three schemes at NLL (upper right), NNLL (lower left), and N$^3$LL (lower right). 
It is striking to see how the difference among the three schemes is still rather large at NLL, then strongly decreases at NNLL,
and finally becomes extremely small at N$^3$LL. Of course, one does expect the details of how the expansions are performed
to matter less and less with increasing logarithmic order. Nevertheless, the level at which the resummed predictions become
independent of the resummation scheme at NNLL and especially at N$^3$LL is truly remarkable. 

\begin{figure}
\vspace*{-5mm}
\includegraphics[scale=0.9]{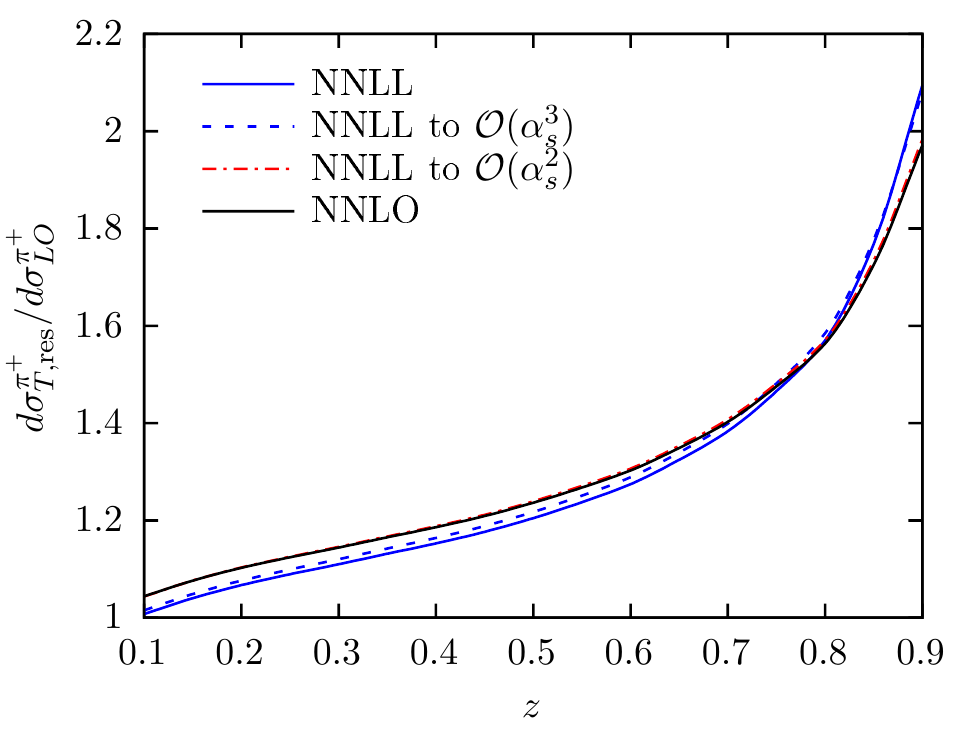}

\vspace*{-6.75cm}
\hspace*{8.6cm}
\includegraphics[scale=0.9]{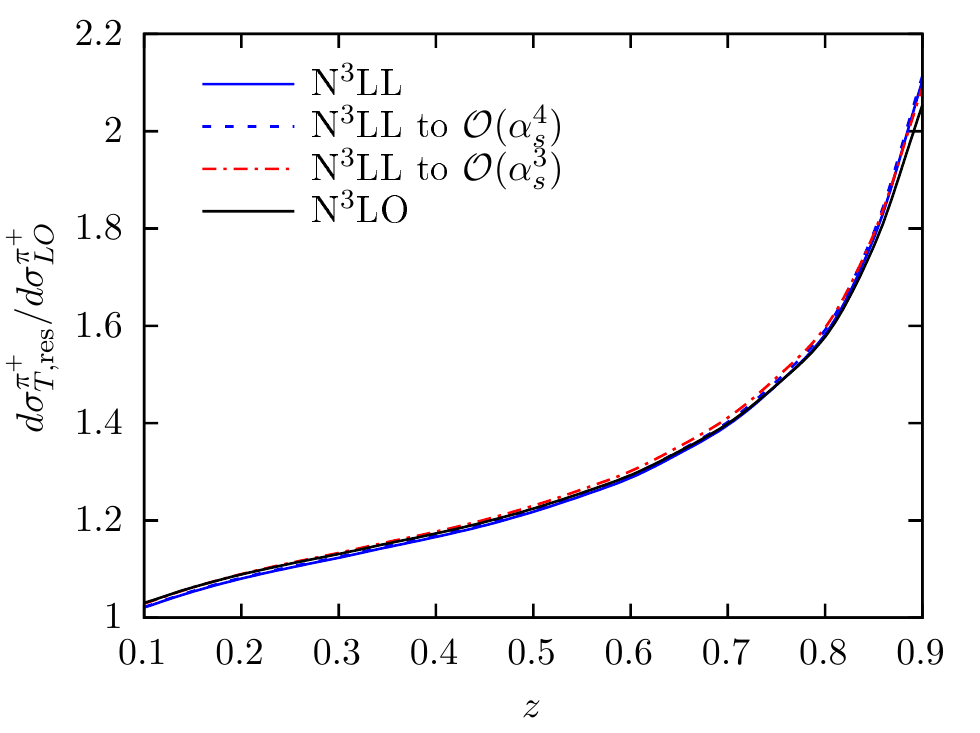}

\caption{{\it Left: NNLO and NNLL-resummed results for the unpolarized $\ell p\to \ell\pi^+X$ transverse cross section in the $q\to q$ channel
normalized to the LO one,
for EIC kinematics with $x\in [0.1,0.8]$. We also show numerical expansions of the NNLL result to ${\cal O}(\alpha_s^2)$ and ${\cal O}(\alpha_s^3)$. 
Right: Same as left, but for $N^{3}$LL, $N^{3}$LO and expansions to ${\cal O}(\alpha_s^3)$ and ${\cal O}(\alpha_s^4)$.}
\label{fig2}}
\end{figure}

Encouraged by these observations, we now turn to fixed-order expansions of our resummed results. Figure~\ref{fig2} (left) shows
again the NNLL-resummed result for scheme (a), along with its expansion to NNLO as given by Eqs.~(\ref{eqnlo1}) and~(\ref{eqnlo1a})
(black solid line) and already obtained in Ref.~\cite{Abele:2021nyo}. All results are again normalized to the LO cross section. 
 We note that finite-order expansions are independent of the resummation scheme chosen.
We observe that resummation within scheme (a) leads to a suppression of the cross section at lower $z$ and to the
expected enhancement at high $z$ where the threshold logarithms become particularly important. In addition to these
two results, we also expand the resummed cross section numerically to orders $\alpha_s^2$ and $\alpha_s^3$. As expected, the result 
for the ${\cal O}(\alpha_s^2)$ expansion (dash-dotted line) is extremely close to the NNLO one. The only difference between
these two results comes from the fact that the formal expansion of the NLP factors in~(\ref{nlpfactors}) will produce
also terms with higher inverse powers of $N$ and $M$, as noted at the end of Sec.~\ref{N3LOexp}. These terms are not
included in our explicit NNLO expansions, but do contribute to the numerical ${\cal O}(\alpha_s^2)$ expansion of the cross
section. As one can see by comparing the two corresponding curves, they are of very small size. 
The dashed line in the left part of Fig.~\ref{fig2} shows the ${\cal O}(\alpha_s^3)$ expansion of the NNLL resummed cross
section. We observe that this result is already very close to the full NNLL one, indicating that terms of order ${\cal O}(\alpha_s^4)$
or higher are small. 

The right part of Fig.~\ref{fig2} presents the same analysis one order higher. We show the N$^3$LL-resummed cross section for scheme (a),
and now the expansion to N$^3$LO as given by Eqs.~(\ref{eqnlo1}),(\ref{eqnlo1a}) and~(\ref{eqnlo1b}). This time, we numerically
expand the N$^3$LL result to orders $\alpha_s^3$ and $\alpha_s^4$. Again the numerical expansion to ${\cal O}(\alpha_s^3)$  essentially 
coincides with the approximate N$^3$LO one, up to tiny corrections suppressed as $1/N^2,1/M^2$ or higher. The result at
${\cal O}(\alpha_s^3)$ is almost indistinguishable from the full N$^3$LL-resummed one, demonstrating again that corrections 
beyond third order are all but negligible.  We note that the ${\cal O}(\alpha_s^3)$ expansion obtained from the N$^3$LL-resummed
result is more complete than the ${\cal O}(\alpha_s^3)$ expansion shown in the left part of Fig.~\ref{fig2}: It contains all {\it seven}
``towers'' of threshold logarithms, that is, terms of the form $\alpha_s^3 \ln^n(N)\ln^m(M)$ with $0\leq n+m\leq 6$, whereas
NNLL resummation can only correctly reproduce the five towers with $2\leq n+m\leq 6$.

\section{Conclusions and outlook}

We have explored higher-order QCD corrections to the quark-to-quark hard-scattering cross section relevant for semi-inclusive
DIS. We have developed the threshold resummation framework for SIDIS to N$^3$LL accuracy, hereby extending previous work
carried out at NNLL~\cite{Abele:2021nyo}. Among the main tasks to be completed for achieving N$^3$LL resummation was 
the derivation of the three-loop hard factor from the spacelike form factor. We have used our N$^3$LL results to derive
approximate N$^3$LO corrections for SIDIS. These corrections contain all seven ``towers'' of threshold logarithms
that are present at this order. We have also included dominant subleading logarithmic terms that are
suppressed near threshold. 

We have presented phenomenological results for resummed and approximate fixed-order SIDIS cross sections for
EIC kinematics. These show an excellent perturbative stability of the cross section in the sense that the N$^3$LL cross
section is only modestly enhanced over the NNLL one, and that generally corrections beyond ${\cal O}(\alpha_s^3)$ 
seem unimportant. A particularly striking result is that the actual treatment of resummation, in terms of how
the relevant expansions are carried out in practice, matters less and less when the logarithmic accuracy of resummation
increases, so that the N$^3$LL result is essentially insensitive to the resummation scheme adopted. 
Clearly, our results show that the SIDIS cross section may serve as an excellent testbed for studies of 
higher orders in perturbation theory. We believe that our results are a valuable addition to the general ``library'' of 
QCD observables that are known to NNLO and beyond. 

Future extensions of this work should also address non-perturbative power corrections to the SIDIS cross section,
very little about which is currently known. It would be an interesting phenomenlogical study to confront experimental
data with our perturbative results at various high orders ranging from NLO to N$^3$LL, ascertaining how
the size of phenomenlogically extracted power corrections depends on the order of perturbation theory that is employed.

We finally note that while we have focused our studies entirely on the spin-averaged SIDIS cross section,
all our results equally apply to the helicity-dependent one. More precisely, the N$^3$LL result and hence its
approximate N$^3$LO expansion are identical in the spin-averaged and the spin-dependent cases. This further
corroborates the finding of Ref.~\cite{Anderle:2013lka} that the SIDIS spin asymmetry is insensitive to 
higher-order perturbative QCD corrections.

\section*{Acknowledgements}

This study was supported in part by Deutsche
Forschungsgemeinschaft (DFG) through the Research Unit
FOR 2926 (Project No. 40824754). 

\appendix
\section{Appendix: Coefficients for resummation to N$^3$LL}
\label{sec:appendixAndim}

We use the following expansion of the running strong coupling~\cite{Vogt:2000ci,Catani:2003zt}
\beeq
\alpha_{s}(\mu)
	&=&\frac{\alpha_{s}(\mu_{R}) }{X}
		-\left(\frac{\alpha_{s}(\mu_{R})}{X}\right)^{2} \frac{b_{1}}{b_{0}} \ln X\nn\\[2mm]
	&+&\left(\frac{\alpha_{s}(\mu_{R})}{X}\right)^{3} \left(\frac{b_{1}^2}{b_{0}^2 } \left(\ln ^2 X-\ln X+X-1\right)-\frac{b_{2}}{b_{0}} (X-1)\right) \nn \\[2mm]
	&+& \left(\frac{\alpha_{s}(\mu_{R})}{X}\right)^{4}\left( \frac{b_1^3}{b_0^3} \left(X-\frac{1}{2}X^2-\ln ^3 X+\frac{5}{2} \ln ^2 X+2 (1-X) \ln X-\frac{1}{2}\right) \right. \nn \\[2mm]
	&& \qquad\;\, +\left.\frac{b_1 b_2 }{b_0^2}(-X (1-X)+2 X \ln X-3 \ln X)+\frac{b_3}{2 b_0} \left(1-X^2\right) \right)\,,
\label{eq:asmu}
\eeeq
where
\begin{equation}
X\,\equiv \,1 + b_{0} \alpha_{s} (\mu_{R}) \ln \frac{\mu^{2}}{\mu^{2}_{R}}\,,
\end{equation}
and
\bea
b_0 & = & \f{1}{12\pi} \left(11C_A-2 N_f\right)\; , \qquad b_1 = \f{1}{24\pi^2}\left(17C_A^2-5C_AN_f-3C_F N_f\right) \; , \nn \\[2mm]
b_2 & = & \f{1}{64\pi^3}\left(\f{2857}{54} C_A^3- \f{1415}{54} C_A^2 N_f-\f{205}{18} C_A C_F N_f+C_F^2 N_f+
\f{79}{54} C_A N_f^2+\f{11}{9} C_F N_f^2\right) ,\;\nn \\[2mm]
b_3 &=& \frac{1}{256 \pi^4}\left[\left( \frac{149753}{6} + 3564 \zeta(3) \right)
        - \left( \frac{1078361}{162} + \frac{6508}{27} \zeta(3) \right) N_f
  \right.\nonumber \\ & &
      \left. + \left( \frac{50065}{162} + \frac{6472}{81} \zeta(3) \right) N_f^2
       +  \frac{1093}{729}  N_f^3 \right]\,,
\eea
with $N_f$ the number of flavors and 
\beq\label{cqcg} 
C_F\,=\,\frac{N_c^2-1}{2N_c}\,=\,\frac{4}{3}  
\;, \;\;\;C_A\,=\,N_c=3 \; .
\eeq
For the $b_3$ coefficient we have inserted the values of $C_F,C_A$; the full expression can be found in~\cite{vanRitbergen:1997va}.
The relevant expansion coefficients for $A_q$ in Eq.~(\ref{Qexp}) read~\cite{Vogt:2000ci,KT,Moch:2004pa,Harlander:2001is,eric,Moch:2018wjh,vonManteuffel:2020vjv,Henn:2019swt}:
\beeq 
\label{A12coef} 
A_q^{(1)} &=&C_F
\;,\;\;\;\;\quad A_q^{(2)}\;=\;\frac{1}{2} \; C_F \left[ 
C_A \left( \frac{67}{18} - \frac{\pi^2}{6} \right)  
- \frac{5}{9} N_f \right] \; , \nn \\[2mm] 
A_q^{(3)}&=&\f{1}{4}C_F\left[C_A^2 \left(\f{245}{24}-\f{67}{9}\zeta(2)+
\f{11}{6}\zeta(3)+\f{11}{5}\zeta(2)^2 \right)+C_F N_f\left(-\f{55}{24}+2\zeta(3) \right)\right. \nn \\[2mm]
&& \qquad\;\, \left. +C_A N_f\left(-\f{209}{108}+\f{10}{9}\zeta(2)-\f{7}{3}\zeta(3) \right) -\f{1}{27}N_f^2 \right] \nn\\[2mm]
A_q^{(4)} &=& C_F\bigg[
  C_A^3 \left( \frac{1309 \zeta_{3}}{432}-\frac{11 \pi ^2 \zeta_{3}}{144}-\frac{\zeta_{3}^2}{16}-\frac{451 \zeta_{5}}{288}+\frac{42139}{10368}-\frac{5525 \pi^2}{7776}+\frac{451 \pi ^4}{5760}-\frac{313 \pi ^6}{90720} \right) 
  \nn\\[2mm]&&
+ N_f T_F  C_A^2 \left( -\frac{361 \zeta_{3}}{54}+\frac{7 \pi ^2 \zeta_{3}}{36}+\frac{131 \zeta_{5}}{72} -\frac{24137}{10368}+\frac{635 \pi ^2}{1944}-\frac{11 \pi ^4}{2160}  \right)  
  \nn\\[2mm]&&
 +  N_f T_F   C_F C_A \left(  \frac{29 \zeta_{3}}{9}-\frac{\pi ^2 \zeta_{3}}{6}+\frac{5 \zeta_{5}}{4}-\frac{17033}{5184}+\frac{55 \pi ^2}{288}-\frac{11 \pi^4}{720} \right)  
  \nn\\&&
 + N_f T_F  C_F^2 \left( \frac{37 \zeta _3}{24}-\frac{5 \zeta _5}{2}+\frac{143}{288} \right) \nn
 + (N_f T_F)^2   C_A \left( \frac{35 \zeta _3}{27}-\frac{7 \pi ^4}{1080}-\frac{19 \pi ^2}{972}+\frac{923}{5184} \right)    
  \nn\\[2mm]&&
+ (N_f T_F)^2   C_F \left(-\frac{10 \zeta _3}{9}+\frac{\pi ^4}{180}+\frac{299}{648}\right) 
+ (N_f T_F)^3  \left(-\frac{1}{81}+\frac{2 \zeta _3}{27}\right)
 \bigg]  \nn
 \\[2mm]
  &&+
  \frac{d^{abcd}_F d^{abcd}_A}{N_c} \left(  \frac{\zeta_{3}}{6}-\frac{3 \zeta_{3}^2}{2}+\frac{55 \zeta_{5}}{12}-\frac{\pi^2}{12}-\frac{31 \pi ^6}{7560}   \right)
  +N_f \frac{d^{abcd}_F d^{abcd}_F}{N_c} \left( \frac{\pi^2}{6}-\frac{\zeta_3}{3}-\frac{5\zeta_5}{3} \right) 
\,,  \;\;\;
\eeeq
with
 \beeq 
 \frac{d_{F}^{abcd}d_{A}^{abcd}}{N_{c}^2-1} =\frac{N_c (N_c^2+6)}{48} ,\,\,\,\,\,\,\,  \frac{d_{F}^{abcd}d_{F}^{abcd}}{N_{c}^2-1} = \frac{N_c^4-6 N_c^2+18}{96 N_c^2},\,\,\,\,\,\,\, 
 T_{F}& = \frac{1}{2} \, .
\eeeq
Furthermore for the expansion of the function $\widehat{D}_q$ we have~\cite{Vogt:2000ci,Catani:2001ic,Catani:2003zt,Catani:2014uta,Hinderer:2018nkb},
\bea\label{eq:Dhat}
\widehat{D}_q^{(2)}
&=&C_F\,\left[C_A\left(-\f{101}{27}+\f{7}{2}\,\zeta(3)\right) + \f{14}{27}\,N_f   \right]\, , \nn\\[2mm]
\widehat{D}_q^{(3)}
&=& C_{F}\left[C^{2}_{A}\left(-\frac{1}{36} 11 \pi ^2 \zeta (3)+\frac{1541 \zeta (3)}{108}-6 \zeta (5)+\frac{11 \pi ^4}{720}+\frac{799 \pi ^2}{1944}-\frac{297029}{23328}\right)\right.\nn\\[2mm]
&& \qquad\;\, +C_{A}N_{f}\left(-\frac{113 \zeta (3)}{108}+\frac{\pi ^4}{360}-\frac{103 \pi ^2}{1944}+\frac{31313}{11664}\right) + N^{2}_{f}\left( -\frac{\zeta (3)}{9}-\frac{58}{729} \right) \nn\\[2mm]
&& \qquad\;\, \left. +C_{F} N_{f}\left(-\frac{19 \zeta (3)}{18}-\frac{\pi ^4}{180}+\frac{1711}{864}\right)\right]\, .
\eea
The expansion coefficients for $\widehat{C}_{qq}$ in Eq.~(\ref{SIDISres}) read~\cite{Catani:2003zt}
\beeq
\widehat{C}_{qq}^{(1)}
	&=& \frac{\pi ^2 }{3}A^{(1)}_{q}\,,\nn\\[2mm]
\widehat{C}_{qq}^{(2)}
	&=&\frac{\pi ^4 }{18}\big(A^{(1)}_q\big)^2
		+A^{(1)}_{q} \pi b_{0} \left(\frac{\pi ^2}{3}  \ln \frac{\mu^{2}_{R}}{Q^{2}} 
		+\frac{8}{3} \zeta (3)  \right) 
		+\frac{\pi ^2 }{3}A^{(2)}_{q}\,,\nn \\[2mm]
\widehat{C}_{qq}^{(3)}
	&=& \frac{\pi^6}{162}\big(A^{(1)}_q\big)^3  -\frac{1}{3} \pi ^3 b_0\widehat{D}_q^{(2)} + \frac{1}{9}\big(A^{(1)}_q\big)^2 b_0 \pi ^3  \left(\pi ^2 \ln \frac{\mu^{2}_{R}}{Q^{2}}+8 \zeta (3)\right) \nn\\[2mm]
	&+& A^{(1)}_{q} b_1\left(\frac{\pi ^4}{3}\ln\frac{\mu^{2}_{R}}{Q^{2}}+\frac{8 \pi ^2 \zeta (3)}{3}\right) + A^{(1)}_{q} b^2_0\left(\frac{\pi ^4 }{3}\ln^2 \frac{\mu^{2}_{R}}{Q^{2}}+\frac{16}{3} \pi ^2\zeta (3) \ln \frac{\mu^{2}_{R}}{Q^{2}} -\frac{\pi ^6}{45}\right)\nn\\[2mm]
	&+&  \frac{2}{3}A^{(2)}_{q}  \pi  b_0 \left(\pi ^2 \ln \frac{\mu^{2}_{R}}{Q^{2}}+8 \zeta (3)\right) +\frac{\pi ^4}{9}A^{(1)}_{q}A^{(2)}_{q} +\frac{\pi ^2}{3} A^{(3)}_{q}\,.
\eeeq
Finally, for $H_{qq}^{\mathrm{SIDIS}}$ we find for an arbitrary renormalization scale $\mu_R$, but for $\mu_F=Q$: 
\beeq\label{H12sidis}
H_{qq}^{\mathrm{SIDIS},(1)} 
	&=& C_{F}\left(-4-\frac{\pi ^2}{6}\right)\,,\nn \\[2mm]
H_{qq}^{\mathrm{SIDIS},(2)} 
	&= & C_{F}\left(-4-\frac{\pi ^2}{6}\right)\pi b_{0}\ln \frac{\mu^{2}_{R}}{Q^{2}}
		+C^{2}_{F}\left(-\frac{15  \zeta (3)}{4}+\frac{61 \pi ^2 }{48}+\frac{511}{64}-\frac{\pi ^4 }{60}\right)\nn\\[2mm]
	&+&C_{F}C_{A}\left(\frac{7 \zeta (3)}{4}+\frac{3 \pi ^4 }{80}-\frac{1535 }{192}-\frac{403 \pi ^2 }{432} \right)
		+C_{F} N_{f}\left(\frac{ \zeta (3)}{2}+\frac{29 \pi ^2}{216}+\frac{127 }{96} \right)\,,
\eeeq
and, in Sec.~\ref{hardfactor}, the three-loop contribution
\beeq
H_{qq}^{\mathrm{SIDIS},(3)}
	&=& C^{3}_{F} \left( \frac{\zeta (3)^2}{2}+\frac{25 \pi ^2 \zeta (3)}{12}-\frac{115 \zeta (3)}{16}+\frac{83 \zeta (5)}{4} \right. \nn \\[2mm]
	&& \left. \qquad\quad	+\frac{1937 \pi ^6}{136080}-\frac{5599}{384}-\frac{181 \pi ^4}{960}-\frac{4729 \pi ^2}{1152} \right)\nn\\[2mm]
	&+& C^{2}_{F}C_{A}\left( \frac{37 \zeta (3)^2}{12}-\frac{571 \pi ^2 \zeta (3)}{216}-\frac{8653 \zeta (3)}{432}-\frac{689 \zeta (5)}{72} \right. \nn \\[2mm]
	&& \left. \qquad\qquad	+\frac{2603 \pi ^4}{38880}+\frac{93581 \pi ^2}{10368}+\frac{74321}{2304}-\frac{227 \pi ^6}{17010} \right) \nn\\[2mm]
	&+& C_{F} C^{2}_{A}\left( -\frac{25 \zeta (3)^2}{12}+\frac{571 \pi ^2 \zeta (3)}{288}+\frac{82385 \zeta (3)}{5184}-\frac{51 \zeta (5)}{16} \right. \nn \\[2mm]
	&& \left. \qquad\qquad	+\frac{41071 \pi ^4}{311040}-\frac{51967 \pi ^2}{10368}-\frac{125 \pi ^6}{27216}-\frac{1505881}{62208} \right)\nn\\[2mm]
	&+&C^{2}_{F} N_{f}\left( -\frac{1}{27} 7 \pi ^2 \zeta (3)+\frac{869 \zeta (3)}{216}-\frac{19 \zeta (5)}{18}-\frac{421}{192}-\frac{1363 \pi ^2}{1296}-\frac{157 \pi ^4}{4860} \right) \nn\\[2mm]
	&+& C_{F} C_{A} N_{f}\left( -\frac{1}{72} 5 \pi ^2 \zeta (3)-\frac{94 \zeta (3)}{81}-\frac{\zeta (5)}{8}+\frac{10595 \pi ^2}{7776}+\frac{110651}{15552}-\frac{1259 \pi ^4}{77760} \right) \nn\\[2mm]
	&+& C_{F} N^{2}_{f} \left( -\frac{79 \zeta (3)}{324}-\frac{\pi ^4}{3888}-\frac{307 \pi ^2}{3888}-\frac{7081}{15552}\right)\nn\\[2mm]
	&+& C_{F} N_{f,V} \left(\frac{C_{A}^{2}-4}{C_{A}}\right)\left( \frac{7 \zeta (3)}{48}-\frac{5 \zeta (5)}{6}+\frac{5 \pi ^2}{96}+\frac{1}{8}-\frac{\pi ^4}{2880} \right)\nn \\[2mm]
	&+& \left[C_{F} C_{A}  \left( \frac{7 \zeta (3)}{2}+\frac{3 \pi ^4}{40}-\frac{1535}{96}-\frac{403 \pi ^2}{216} \right) + C_{F} N_{f} \left( \zeta (3)+\frac{29 \pi ^2}{108}+\frac{127}{48} \right) \right. \nn \\[2mm]
	&+& \left. C_F C_A N_f\left(\frac{\zeta (3)}{3}+\frac{767 \pi ^2}{1296}-\frac{\pi ^4}{80}+\frac{853}{144}\right) \right.\nn\\[2mm]
	&+&\left. C^{2}_{F}\left( -\frac{15 \zeta (3)}{2}+\frac{61 \pi ^2}{24}-\frac{\pi ^4}{30}+\frac{511}{32} \right)\right]\pi b_{0}\ln \frac{\mu^{2}_{R}}{Q^{2}} \nn\\[2mm]
	&+& C_{F}\left(-4-\frac{\pi ^2}{6}\right) \pi^{2} b_{1}\ln\frac{\mu^{2}_{R}}{Q^{2}} +  C_{F}\left(-4-\frac{\pi ^2}{6}\right) \pi^{2}b^{2}_{0}\ln^{2}\frac{\mu^{2}_{R}}{Q^{2}}\,.
\eeeq

\section{Appendix: NLO and NNLO hard-scattering coefficient functions \label{appendix:hardscat}}

The full NLO coefficient function can be found in~\cite{deFlorian:1997zj,Anderle:2012rq}. Its near-threshold approximation was given 
in~\cite{Abele:2021nyo} and reads
\beeq\label{eqnlo1}
\tilde{\omega}^{T,(1)}_{qq}\left(N,M, \frac{\mu_{R}}{Q}, \frac{\mu_{F}}{Q}\right)&=&
e^{2}_{q}C_{F}\left\{2{\cal L}^2+\frac{\pi ^2}{6}-4+\left(-\frac{3}{2}+2{\cal L}\right)\ln \frac{\mu_{F}^{2}}{Q^{2}}+ 
{\cal L}\left(\frac{1}{N}+\frac{1}{M}\right)\right\}\,,\quad
\eeeq
where ${\cal L}\,\equiv\,\frac{1}{2}\left( \ln (\bar{N})  + \ln (\bar{M})\right)$. The corresponding approximate NNLO result is given by
\beeq \label{eqnlo1a}
&&\hspace*{-1.5cm}\frac{1}{e_q^2}\,\tilde{\omega}^{T,(2)}_{qq}\left(N,M, \frac{\mu_{R}}{Q}, \frac{\mu_{F}}{Q}\right)\,=\,
2C_F^2 {\cal L}^4 + 4C_F {\cal L}^3 \left(\frac{\pi}{3} \,b_0+ C_F\ln\frac{\mu^{2}_{F}}{Q^{2}}\right)\nn\\[2mm]
&+&\,C_F{\cal L}^2\left[ C_F \left(-8+\frac{\pi^2}{3}+2 \ln^2\frac{\mu^{2}_{F}}{Q^{2}}-
3 \ln\frac{\mu^{2}_{F}}{Q^{2}}\right)+\left(\frac{67}{18}-\frac{\pi ^2}{6}\right) C_{A}-\frac{5}{9} N_{f}\right]\nn\\[2mm]
&+&\,C_F {\cal L}\left[ \left(\frac{101}{27}-\frac{7}{2}\,\zeta(3)\right)\,C_A -\frac{14}{27}\,N_f+
C_F \ln\frac{\mu^{2}_{F}}{Q^{2}} \left( -8+\frac{\pi^2}{3}-3\ln\frac{\mu^{2}_{F}}{Q^{2}}\right)\right.\nn\\[2mm]
&&\left.\,+\,\left( \left(\frac{67}{18}-\frac{\pi ^2}{6}\right) C_{A}-\frac{5}{9} N_{f}\right)  \ln\frac{\mu^{2}_{F}}{Q^{2}}
-\pi b_0 \ln^2\frac{\mu^{2}_{F}}{Q^{2}}\right]\nn\\[2mm]
&+&\,C_F^2\left[ \frac{511}{64}-\frac{\pi^2}{16}-\frac{\pi^4}{60}-\frac{15}{4}\,\zeta(3)+
\ln\frac{\mu^{2}_{F}}{Q^{2}}\left( \frac{9}{8}\ln\frac{\mu^{2}_{F}}{Q^{2}}+\frac{93}{16}-3 \zeta(3)\right)\right]  \nn\\[2mm]
&+&\,C_F C_A \left[ -\frac{1535}{192}-\frac{5\pi^2}{16}+\frac{7\pi^4}{720}+\frac{151}{36}\,\zeta(3)\right]+
C_F N_f\left[ \frac{127}{96}+\frac{\pi^2}{24}+\frac{\zeta(3)}{18}\right]  \nn\\[2mm]
&+&\,\frac{3}{4}\,C_F \pi b_0 \ln^2\frac{\mu^{2}_{F}}{Q^{2}}-\frac{C_F \pi^3 b_0}{3}\ln\frac{\mu^{2}_{F}}{Q^{2}}+C_F \left(
-\frac{17}{48}\,C_A + \frac{3}{2}\,\zeta(3)C_A+\frac{N_f}{24}\right)\ln\frac{\mu^{2}_{F}}{Q^{2}}\nn\\[2mm]
&+&\,\pi b_0 \ln\frac{\mu^{2}_{R}}{Q^{2}}\;\frac{1}{e_q^2}\,\tilde{\omega}^{T,(1)}_{qq}\left(N,M, \frac{\mu_{R}}{Q}, \frac{\mu_{F}}{Q}\right)\,
+\,2\,C_F^2\, {\cal L}^3 \left(\frac{1}{N}+\frac{1}{M}\right)\,.
\eeeq

%%%%%%%%%%%%%%%%%%%%%%%%%%%%%%%%%%%%%%%%%%%%%%%%%%%%%%%%
%%%%%%%%%%%%%%%%%%%%%%%%%%%%%%%%%    References    %%%%%%%%%%%%%%%
%%%%%%%%%%%%%%%%%%%%%%%%%%%%%%%%%%%%%%%%%%%%%%%%%%%%%%%%

%\vspace*{-3mm}


\newpage
\begin{thebibliography}{99}

\bibitem{deFlorian:2009vb}
D.~de Florian, R.~Sassot, M.~Stratmann and W.~Vogelsang,
%``Extraction of Spin-Dependent Parton Densities and Their Uncertainties,''
Phys. Rev. D \textbf{80}, 034030 (2009)
%doi:10.1103/PhysRevD.80.034030
[arXiv:0904.3821 [hep-ph]].

\bibitem{Leader:2010rb}
E.~Leader, A.~V.~Sidorov and D.~B.~Stamenov,
%``Determination of Polarized PDFs from a QCD Analysis of Inclusive and Semi-inclusive Deep Inelastic Scattering Data,''
Phys. Rev. D \textbf{82}, 114018 (2010)
%doi:10.1103/PhysRevD.82.114018
[arXiv:1010.0574 [hep-ph]].

\bibitem{deFlorian:2014xna}
D.~de Florian, R.~Sassot, M.~Epele, R.~J.~Hern\'andez-Pinto and M.~Stratmann,
%``Parton-to-Pion Fragmentation Reloaded,''
Phys. Rev. D \textbf{91}, no.1, 014035 (2015)
%doi:10.1103/PhysRevD.91.014035
[arXiv:1410.6027 [hep-ph]].

\bibitem{Ethier:2017zbq}
J.~J.~Ethier, N.~Sato and W.~Melnitchouk,
%``First simultaneous extraction of spin-dependent parton distributions and fragmentation functions from a global QCD analysis,''
Phys. Rev. Lett. \textbf{119}, no.13, 132001 (2017)
doi:10.1103/PhysRevLett.119.132001
[arXiv:1705.05889 [hep-ph]].

\bibitem{Bertone:2017tyb}
V.~Bertone \textit{et al.} [NNPDF],
%``A determination of the fragmentation functions of pions, kaons, and protons with faithful uncertainties,''
Eur. Phys. J. C \textbf{77}, no.8, 516 (2017)
%doi:10.1140/epjc/s10052-017-5088-y
[arXiv:1706.07049 [hep-ph]].

\bibitem{Moffat:2021dji}
E.~Moffat \textit{et al.} [Jefferson Lab Angular Momentum (JAM)],
%``Simultaneous Monte~Carlo analysis of parton densities and fragmentation functions,''
Phys. Rev. D \textbf{104}, no.1, 016015 (2021)
doi:10.1103/PhysRevD.104.016015
[arXiv:2101.04664 [hep-ph]].

\bibitem{Khalek:2021gxf}
R.~A.~Khalek, V.~Bertone and E.~R.~Nocera,
%``Determination of unpolarized pion fragmentation functions using semi-inclusive deep-inelastic-scattering data,''
Phys. Rev. D \textbf{104}, no.3, 034007 (2021)
%doi:10.1103/PhysRevD.104.034007
[arXiv:2105.08725 [hep-ph]].

\bibitem{Abdolmaleki:2021yjf}
H.~Abdolmaleki \textit{et al.} [xfitter Developers\textquoteright{} Team],
%``QCD analysis of pion fragmentation functions in the xFitter framework,''
Phys. Rev. D \textbf{104}, no.5, 056019 (2021)
doi:10.1103/PhysRevD.104.056019
[arXiv:2105.11306 [hep-ph]].

\bibitem{Borsa:2021ran}
I.~Borsa, D.~de Florian, R.~Sassot and M.~Stratmann,
%``Pion fragmentation functions at high energy colliders,''
Phys. Rev. D \textbf{105}, no.3, L031502 (2022)
%doi:10.1103/PhysRevD.105.L031502
[arXiv:2110.14015 [hep-ph]].

\bibitem{Borsa:2022vvp}
I.~Borsa, D.~de Florian, R.~Sassot, M.~Stratmann and W.~Vogelsang,
%``Towards a Global QCD Analysis of Fragmentation Functions at Next-To-Next-To-Leading Order Accuracy,''
[arXiv:2202.05060 [hep-ph]].

\bibitem{Aschenauer:2019kzf}
E.~C.~Aschenauer, I.~Borsa, R.~Sassot and C.~Van Hulse,
%``Semi-inclusive Deep-Inelastic Scattering, Parton Distributions and Fragmentation Functions at a Future Electron-Ion Collider,''
Phys. Rev. D \textbf{99}, no.9, 094004 (2019)
%doi:10.1103/PhysRevD.99.094004
[arXiv:1902.10663 [hep-ph]].

%\cite{Abele:2021nyo}
\bibitem{Abele:2021nyo}
M.~Abele, D.~de Florian and W.~Vogelsang,
%``Approximate NNLO QCD corrections to semi-inclusive DIS,''
Phys. Rev. D \textbf{104}, no.9, 094046 (2021)
%doi:10.1103/PhysRevD.104.094046
[arXiv:2109.00847 [hep-ph]].
%6 citations counted in INSPIRE as of 12 Mar 2022

\bibitem{Cacciari:2001cw}
M.~Cacciari and S.~Catani,
%``Soft gluon resummation for the fragmentation of light and heavy quarks at large x,''
Nucl. Phys. B \textbf{617}, 253-290 (2001)
%doi:10.1016/S0550-3213(01)00469-2
[arXiv:hep-ph/0107138 [hep-ph]].

%\cite{Anderle:2012rq}
\bibitem{Anderle:2012rq}
D.~P.~Anderle, F.~Ringer and W.~Vogelsang,
%``QCD resummation for semi-inclusive hadron production processes,''
Phys. Rev. D \textbf{87}, no.3, 034014 (2013)
%doi:10.1103/PhysRevD.87.034014
[arXiv:1212.2099 [hep-ph]].
%34 citations counted in INSPIRE as of 21 Jul 2021

%\cite{Anderle:2013lka}
\bibitem{Anderle:2013lka}
D.~P.~Anderle, F.~Ringer and W.~Vogelsang,
%``Threshold resummation for polarized (semi-)inclusive deep inelastic scattering,''
Phys. Rev. D \textbf{87}, 094021 (2013)
%doi:10.1103/PhysRevD.87.094021
[arXiv:1304.1373 [hep-ph]].
%7 citations counted in INSPIRE as of 21 Jul 2021

\bibitem{deFlorian:1997zj}
D.~de Florian, M.~Stratmann and W.~Vogelsang,
%``QCD analysis of unpolarized and polarized Lambda baryon production in leading and next-to-leading order,''
Phys. Rev. D \textbf{57}, 5811-5824 (1998)
doi:10.1103/PhysRevD.57.5811
[arXiv:hep-ph/9711387 [hep-ph]].

%\cite{Sterman:2006hu}
\bibitem{Sterman:2006hu}
G.~F.~Sterman and W.~Vogelsang,
%``Crossed Threshold Resummation,''
Phys. Rev. D \textbf{74}, 114002 (2006)
%doi:10.1103/PhysRevD.74.114002
[arXiv:hep-ph/0606211 [hep-ph]].
%18 citations counted in INSPIRE as of 21 Jul 2021

%\cite{Catani:2003zt}
\bibitem{Catani:2003zt}
S.~Catani, D.~de Florian, M.~Grazzini and P.~Nason,
%``Soft gluon resummation for Higgs boson production at hadron colliders,''
JHEP \textbf{07}, 028 (2003)
%doi:10.1088/1126-6708/2003/07/028
[arXiv:hep-ph/0306211 [hep-ph]].
%768 citations counted in INSPIRE as of 21 Jul 2021

\bibitem{Hinderer:2018nkb}
P.~Hinderer, F.~Ringer, G.~Sterman and W.~Vogelsang,
%``Threshold Resummation at NNLL for Single-particle Production in Hadronic Collisions,''
Phys. Rev. D \textbf{99}, no.5, 054019 (2019)
%doi:10.1103/PhysRevD.99.054019
[arXiv:1812.00915 [hep-ph]].
%11 citations counted in INSPIRE as of 21 Jul 2021

\bibitem{Moch:2005ba}
S.~Moch, J.~A.~M.~Vermaseren and A.~Vogt,
%``Higher-order corrections in threshold resummation,''
Nucl. Phys. B \textbf{726}, 317-335 (2005)
%doi:10.1016/j.nuclphysb.2005.08.005
[arXiv:hep-ph/0506288 [hep-ph]].
%173 citations counted in INSPIRE as of 25 Oct 2021

%\cite{Catani:2013tia}
\bibitem{Catani:2013tia}
S.~Catani, L.~Cieri, D.~de Florian, G.~Ferrera and M.~Grazzini,
%``Universality of transverse-momentum resummation and hard factors at the NNLO,''
Nucl. Phys. B \textbf{881} (2014), 414-443
%doi:10.1016/j.nuclphysb.2014.02.011
[arXiv:1311.1654 [hep-ph]].
%148 citations counted in INSPIRE as of 22 Feb 2022

%\cite{Catani:2014uta}
\bibitem{Catani:2014uta}
S.~Catani, L.~Cieri, D.~de Florian, G.~Ferrera and M.~Grazzini,
%``Threshold resummation at N$^3$LL accuracy and soft-virtual cross sections at N$^3$LO,''
Nucl. Phys. B \textbf{888}, 75 (2014)
%doi:10.1016/j.nuclphysb.2014.09.012
[arXiv:1405.4827 [hep-ph]].

%\cite{Baikov:2009bg}
\bibitem{Baikov:2009bg}
P.~A.~Baikov, K.~G.~Chetyrkin, A.~V.~Smirnov, V.~A.~Smirnov and M.~Steinhauser,
%``Quark and gluon form factors to three loops,''
Phys. Rev. Lett. \textbf{102} (2009), 212002
[arXiv:0902.3519 [hep-ph]].

%\cite{Gehrmann:2010ue}
\bibitem{Gehrmann:2010ue}
T.~Gehrmann, E.~W.~N.~Glover, T.~Huber, N.~Ikizlerli and C.~Studerus,
%``Calculation of the quark and gluon form factors to three loops in QCD,''
JHEP \textbf{06}, 094 (2010)
doi:10.1007/JHEP06(2010)094
[arXiv:1004.3653 [hep-ph]].
%244 citations counted in INSPIRE as of 24 Jan 2022

%\cite{Lee:2010cga}
\bibitem{Lee:2010cga}
R.~N.~Lee, A.~V.~Smirnov and V.~A.~Smirnov,
%``Analytic Results for Massless Three-Loop Form Factors,''
JHEP \textbf{04} (2010), 020
[arXiv:1001.2887 [hep-ph]].

%\cite{Lee:2022nhh}
\bibitem{Lee:2022nhh}
R.~N.~Lee, A.~von Manteuffel, R.~M.~Schabinger, A.~V.~Smirnov, V.~A.~Smirnov and M.~Steinhauser,
%``Quark and gluon form factors in four-loop QCD,''
[arXiv:2202.04660 [hep-ph]].

%\cite{Catani:1996yz}
\bibitem{Catani:1996yz}
S.~Catani, M.~L.~Mangano, P.~Nason and L.~Trentadue,
%``The Resummation of soft gluons in hadronic collisions,''
Nucl. Phys. B \textbf{478}, 273-310 (1996)
[arXiv:hep-ph/9604351 [hep-ph]].
%486 citations counted in INSPIRE as of 22 Jan 2022

%\cite{Hou:2019efy}
\bibitem{Hou:2019efy}
T.~J.~Hou, J.~Gao, T.~J.~Hobbs, K.~Xie, S.~Dulat, M.~Guzzi, J.~Huston, P.~Nadolsky, J.~Pumplin and C.~Schmidt, \textit{et al.}
%``New CTEQ global analysis of quantum chromodynamics with high-precision data from the LHC,''
Phys. Rev. D \textbf{103}, no.1, 014013 (2021)
%doi:10.1103/PhysRevD.103.014013
[arXiv:1912.10053 [hep-ph]].
%147 citations counted in INSPIRE as of 27 Oct 2021

%\cite{Anderle:2016czy}
\bibitem{Anderle:2016czy}
D.~P.~Anderle, T.~Kaufmann, M.~Stratmann and F.~Ringer,
%``Fragmentation Functions Beyond Fixed Order Accuracy,''
Phys. Rev. D \textbf{95}, no.5, 054003 (2017)
%doi:10.1103/PhysRevD.95.054003
[arXiv:1611.03371 [hep-ph]].
%25 citations counted in INSPIRE as of 27 Oct 2021




\bibitem{Vogt:2000ci}
A.~Vogt,
%``Next-to-next-to-leading logarithmic threshold resummation for deep inelastic scattering and the Drell-Yan process,''
Phys. Lett. B \textbf{497}, 228 (2001)
%doi:10.1016/S0370-2693(00)01344-7
[arXiv:hep-ph/0010146 [hep-ph]].
%\cite{vanRitbergen:1997va}
\bibitem{vanRitbergen:1997va}
T.~van Ritbergen, J.~A.~M.~Vermaseren and S.~A.~Larin,
%``The Four loop beta function in quantum chromodynamics,''
Phys. Lett. B \textbf{400}, 379-384 (1997)
%doi:10.1016/S0370-2693(97)00370-5
[arXiv:hep-ph/9701390 [hep-ph]].
%1148 citations counted in INSPIRE as of 25 Oct 2021

\bibitem{KT} J.~Kodaira and L.~Trentadue, 
%``Summing Soft Emission In QCD,''
Phys.\ Lett.\ B {\bf 112}, 66 (1982); Phys.\ Lett.\ B {\bf 123}, 
335 (1983); S.~Catani, E.~D'Emilio and L.~Trentadue,
%``The gluon form-factor to higher orders: gluon gluon annihilation at small Q-transverse",
Phys.\ Lett.\ B {\bf 211}, 335 (1988).

\bibitem{Moch:2004pa} S.~Moch, J.~A.~M.~Vermaseren and A.~Vogt,
%  ``The Three loop splitting functions in QCD: The Nonsinglet case,''
  Nucl.\ Phys.\ B {\bf 688}, 101 (2004)
  [hep-ph/0403192].
  
  \bibitem{Harlander:2001is} R.~V.~Harlander and W.~B.~Kilgore,
%``Soft and virtual corrections to p p $\to$ H + X at NNLO,''  
Phys.\ Rev.\  D {\bf 64}, 013015 (2001)  [arXiv:hep-ph/0102241].

\bibitem{eric} T.~O.~Eynck, E.~Laenen and L.~Magnea,  
%``Exponentiation of the Drell-Yan cross section near partonic threshold  in the DIS and MS-bar schemes,''  
JHEP {\bf 0306}, 057 (2003)
[arXiv:hep-ph/0305179];
E.~Laenen and L.~Magnea,  
%``Threshold resummation for electroweak annihilation from DIS data,'' 
Phys.\ Lett.\  B {\bf 632}, 270 (2006) [arXiv:hep-ph/0508284].



%\cite{Moch:2018wjh}
\bibitem{Moch:2018wjh}
S.~Moch, B.~Ruijl, T.~Ueda, J.~A.~M.~Vermaseren and A.~Vogt,
%``On quartic colour factors in splitting functions and the gluon cusp anomalous dimension,''
Phys. Lett. B \textbf{782} (2018), 627-632
%doi:10.1016/j.physletb.2018.06.017
[arXiv:1805.09638 [hep-ph]].
%82 citations counted in INSPIRE as of 22 Feb 2022

%\cite{vonManteuffel:2020vjv}
\bibitem{vonManteuffel:2020vjv}
A.~von Manteuffel, E.~Panzer and R.~M.~Schabinger,
%``Cusp and collinear anomalous dimensions in four-loop QCD from form factors,''
Phys. Rev. Lett. \textbf{124} (2020) no.16, 162001
%doi:10.1103/PhysRevLett.124.162001
[arXiv:2002.04617 [hep-ph]].
%61 citations counted in INSPIRE as of 22 Feb 2022

%\cite{Henn:2019swt}
\bibitem{Henn:2019swt}
J.~M.~Henn, G.~P.~Korchemsky and B.~Mistlberger,
%``The full four-loop cusp anomalous dimension in $\mathcal{N}=4$ super Yang-Mills and QCD,''
JHEP \textbf{04} (2020), 018
%doi:10.1007/JHEP04(2020)018
[arXiv:1911.10174 [hep-th]].
%76 citations counted in INSPIRE as of 22 Feb 2022

%\cite{Catani:2001ic}
\bibitem{Catani:2001ic}
S.~Catani, D.~de Florian and M.~Grazzini,
%``Higgs production in hadron collisions: Soft and virtual QCD corrections at NNLO,''
JHEP \textbf{05}, 025 (2001)
%doi:10.1088/1126-6708/2001/05/025
[arXiv:hep-ph/0102227 [hep-ph]].

\end{thebibliography}
\end{document}